\documentclass[%
 reprint,
 amsmath,amssymb,
 aps,
onecolumn,
prfluids,
notitlepage
]{revtex4-2}

\usepackage{graphicx}
\usepackage{dcolumn}
\usepackage{bm}
\usepackage{tikz}

\newcommand{\rey}{$Re_\lambda$}
\newcommand{\re}{$Re_\lambda$~}

\begin{document}


\title{Revisiting turbulence small-scale behavior using \\ velocity gradient triple decomposition}

\author{Rishita Das}
 \email{rishitadas@tamu.edu}
\affiliation{%
 Aerospace Engineering, Texas A$\&$M University
}%

\author{Sharath S. Girimaji}
\affiliation{
 Ocean Engineering, Texas A$\&$M University
}%

\date{\today}

\begin{abstract}

Turbulence small-scale behavior has been commonly investigated in literature by decomposing the velocity-gradient tensor ($A_{ij}$) into the symmetric strain-rate ($S_{ij}$) and anti-symmetric rotation-rate ($W_{ij}$) tensors. To develop further insight, we revisit some of the key studies using a triple decomposition of the velocity-gradient tensor.   The additive triple decomposition formally segregates the contributions of normal-strain-rate ($N_{ij}$), pure-shear ($H_{ij}$) and rigid-body-rotation-rate ($R_{ij}$). The decomposition not only highlights the key role of shear, but it also provides a more accurate account of the influence of normal-strain and pure rotation on important small-scale features. First, the local streamline topology and geometry are described in terms of the three constituent tensors in velocity-gradient invariants’ space. Using DNS data sets of forced isotropic turbulence, the velocity-gradient and pressure field fluctuations are examined at different Reynolds numbers. At all Reynolds numbers, shear contributes the most and rigid-body-rotation the least toward the velocity-gradient magnitude ($A^2$). Especially, shear contribution is dominant in regions of intermittency (high values of $A^2$). It is also shown that the high-degree of enstrophy intermittency reported in literature is due to the shear contribution toward vorticity rather than that of rigid-body-rotation.  The study also provides an explanation for the absence of intermittency of the pressure-Laplacian, despite the strong intermittency of enstrophy and dissipation fields. Overall, it is demonstrated that triple decomposition offers unique and deeper understanding of velocity-gradient behavior in turbulence.

\end{abstract}

\maketitle


\section{\label{sec:intro}Introduction}

The velocity gradient dynamics and small-scale behavior of turbulence have largely been investigated in literature by decomposing the velocity gradient tensor (VGT, $A_{ij}=\partial u_i/\partial x_j$) into symmetric (strain-rate tensor, $S_{ij}$) and anti-symmetric (rotation-rate or vorticity tensor, $W_{ij}$) parts:
\begin{equation}
     A_{ij} = S_{ij} + W_{ij} \;\;\;\; \text{where} \;\;\; 
     S_{ij} = \frac{1}{2}(A_{ij}+A_{ji}) \; ,\;\; W_{ij} = \frac{1}{2}(A_{ij}-A_{ji})
\label{eq:SW}
\end{equation}
This decomposition has led to important insight into small-scale intermittency \cite{sreenivasan1997phenomenology,yeung2018effects,buaria2019extreme}, intense structures in turbulence \cite{sanada1991statistics,hosokawa1997existence,jimenez1998characteristics,moisy2004geometry} and local streamline geometry \cite{ashurst1987alignment,kerr1987histograms,luthi2009expanding}.
However, recent studies \cite{kolavr2007vortex,gao2019rortex,nagata2019triple} have shown that strain-rate and vorticity do not clearly identify the presence of normal-straining and rigid-body-rotation of the fluid.
The presence of shear in both strain-rate and vorticity often obscures our understanding of some of the fundamental phenomena in turbulence.
The purpose of this work is to revisit some of the prominent results of small-scale turbulence segregating the role of normal-strain, shear and pure rotation in fluid motions.
Reinterpretation of the classical results leads to further clarity and deeper insight into velocity gradient behavior in turbulence.

The triple decomposition in this study partitions the local velocity gradients into three elementary transformations - normal strain, rigid-body-rotation and pure shear (Fig.~\ref{fig:NHR}). The normal strain-rate tensor $N_{ij}$ is a diagonal tensor that represents the compression and extension of the fluid element in different directions in a volume-preserving manner. The rigid-body-rotation tensor $R_{ij}$ is an anti-symmetric tensor that represents pure-rotation of the fluid-element by a certain angular velocity. The shear tensor $H_{ij}$ is a lower triangular tensor containing transverse gradients of the velocity components that represent shearing of the fluid element. 

Kol\'{a}\v{r} \cite{kolavr2007vortex} presented a procedure for triple decomposition of the VGT by extracting the pure-shearing motion from the swirling action of vorticity. The method comprises of determination of a so-called basic reference frame among all possible frame rotations, which is computationally very expensive for a three-dimensional flow field. This technique has been used for vortex-structure identification and investigation of internal shear layers in wall-bounded flows \cite{eisma2015interfaces,vsistek2012fluid,maciel2012method}. It has recently been used for investigating regions of strong shearing or rotation and detecting internal shear layer in homogeneous isotropic turbulence at Taylor Reynolds numbers, $Re_\lambda=27$ and $140$ \cite{nagata2019triple}. 
Aside from Kol\'{a}\v{r}'s method, \citet{gao2019rortex} formulated a "Rortex"-based VGT decomposition for locally fluid-rotational points (VGT has complex eigenvalues) in a turbulent flow field. This method \cite{tian2018definitions} of separating the rigid-body-rotation (Rortex) from shear in vorticity is computationally more reasonable and has been employed in several studies \cite{dong2019new,li2019heat,gui2019analysis,arun2019topology} for investigation of coherent vortex structures in turbulent flows.

The goal of this study is to examine velocity gradient statistics in turbulence using the decomposition of VGT into normal-strain-rate, rigid-body-rotation and pure-shear tensors. We revisit certain important velocity gradient behavior and characterize them in terms of $N_{ij}$,$R_{ij}$ and ${H}_{ij}$. The primary objectives of this study are as follows:
\begin{enumerate}
    \item Develop a triple decomposition strategy valid for the entire turbulent flow field by combining different proposals in literature and derive important kinematic characteristics of the various constituents.
    \item Characterize the local streamline shapes associated with normal-strain, pure-shear and rigid-body-rotation tensors in the phase space of VGT invariants.
    \item Establish the contribution of different velocity gradient constituents in a turbulent flow field as a function of Reynolds number. Recall that the average strain-rate and vorticity contributions are approximately equal in an isotropic flow field.
    \item Examine the velocity gradient constituents conditioned on magnitude at high Reynolds numbers to gain insight into intermittency. It is generally believed that enstrophy is more intermittent than dissipation \cite{yeung2018effects,buaria2019extreme}.
    \item Analyze the behavior of the pressure field as a function of the velocity gradient composition. 
\end{enumerate}

The next section of this work outlines the procedure for triple decomposition of VGT, followed by a comprehensive description of the properties of its constituents and its implication in local streamline geometry. 
In the third section, details about the DNS datasets of forced isotropic turbulence are briefly discussed. The results are illustrated in the fourth section -- velocity gradient composition of a turbulent flow field is examined in detail along with its Reynolds number dependence, followed by investigation of the pressure field conditioned on velocity gradient constituents in high Reynolds number turbulence.
Finally, the important findings of this study are summarized in the conclusions section.

\section{\label{sec:Trip} Triple decomposition of VGT}

\begin{figure*}
\begin{tikzpicture}
\node[above right] (img) at (0,0) {\includegraphics[width=0.3\textwidth]{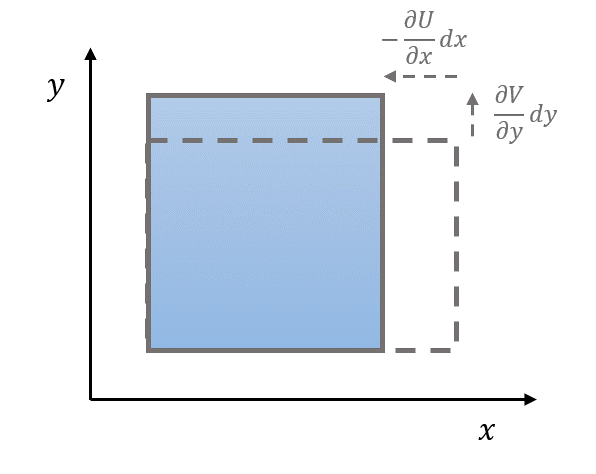}};
\node at (10pt,120pt) {(\textit{a})};
\end{tikzpicture}
\begin{tikzpicture}
\node[above right] (img) at (0,0) {\includegraphics[width=0.3\textwidth]{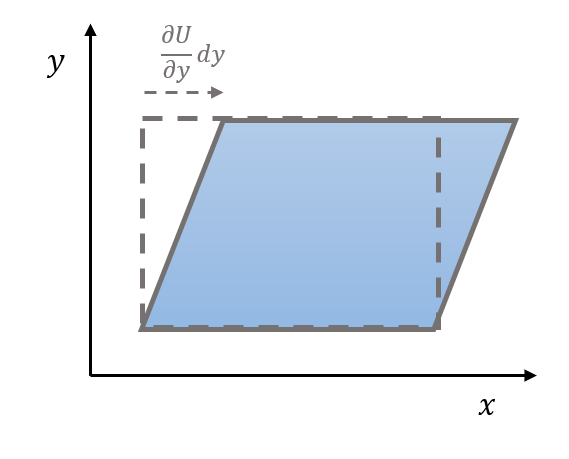}};
\node at (10pt,120pt) {(\textit{b})};
\end{tikzpicture}
\begin{tikzpicture}
\node[above right] (img) at (0,0) {\includegraphics[width=0.3\textwidth]{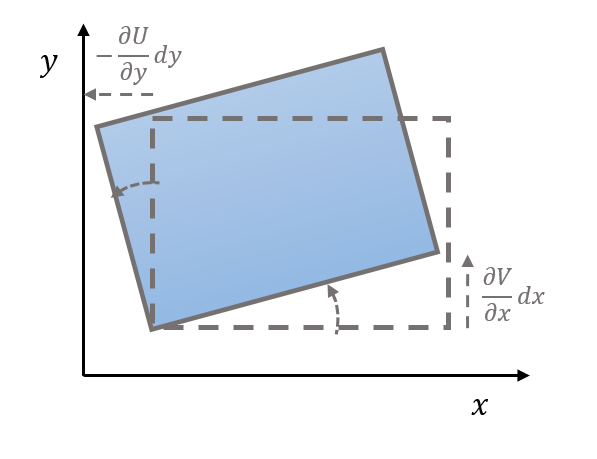}};
\node at (10pt,120pt) {(\textit{c})};
\end{tikzpicture}
\caption{\label{fig:NHR}Two-dimensional example of fluid element deformation due to (a) normal-strain-rate tensor, (b) shear tensor, and (c) rigid-body-rotation tensor}
\end{figure*}
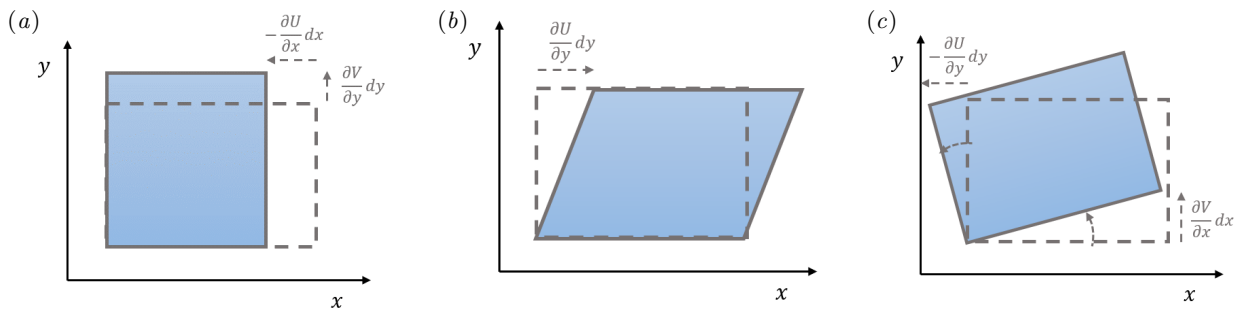

The additive decomposition of the VGT ($\bm{A}$) into normal-strain-rate tensor ($\bm{N}$), rigid-body-rotation-rate tensor ($\bm{R}$) and pure-shear tensor ($\bm{H}$) is given by
\begin{equation}
    A_{ij} = N_{ij} + H_{ij} + R_{ij}
    \label{eq:VGT1}
\end{equation}
The $N_{ij}$, $H_{ij}$ and $R_{ij}$ tensors represent normal-straining, pure-shearing and rigid-body-rotation of a fluid element, respectively.
These transformations are illustrated with some elementary examples in Fig.~\ref{fig:NHR} for reference. 
The decomposition entails considerable level of effort and the technique is different for a fluid element undergoing rigid-body-rotation ($\bm{R}\neq 0$) and one that has no rotation component ($\bm{R} = 0$).
The former is called local fluid rotational while the latter is referred to as local fluid non-rotational \cite{tian2018definitions} in the rest of this work.
In this section, we first outline the decomposition procedure for the two cases. Then we proceed to establish the kinematic properties of the constituent tensors, including the geometric shapes of local streamlines corresponding to each tensor.

\subsection{\label{sec:Trip1} Decomposition procedure}

If $\bm{A}$ has two complex conjugate eigenvalues ($\lambda_{cr}\pm i\lambda_{ci}$) and one real eigenvalue ($\lambda_r$), then it represents locally rotational flow. 
On the other hand, if $\bm{A}$ has only real eigenvalues ($\lambda_1,\lambda_2,\lambda_3$), it implies that the flow is locally non-rotational.
Our procedure combines two different proposals in literature for rotational and non-rotational parts of the flow field.
The procedure of VGT decomposition for both the cases are now discussed in detail.

\subsubsection{\label{sec:Trip1a} Rotational case}

For the rotational case, we follow the VGT decomposition procedure outlined by \citet{gao2019rortex}. This method involves two coordinate-frame rotations to obtain the VGT in the desired lower block triangular form for decomposition. The steps are listed below:
\begin{enumerate}
    \item First we identify the rotational axis ($\bm{\vec{r}}$), which is the real eigenvector of $\bm{A}$. Then the coordinate frame is rotated such that the Z-axis of the new frame is aligned with $\bm{\vec{r}}$. The VGT in this new coordinate frame is given by
    \begin{equation}
        \bm{A'} = \bm{Q} \bm{A} \bm{Q}^T
        \label{eq:rot1}
    \end{equation}
    where, $\bm{Q}$ is a proper rotation matrix obtained from real Schur decomposition of $\bm{A}$ as shown by \citet{liu2018rortex}. 
    \item Next, the coordinate frame is further rotated about the Z-axis, i.e. in the $XY$-plane, by an azimuthal angle $\theta$. This angle $\theta$ is chosen such that the angular velocity of the fluid is minimum. The VGT in this new coordinate frame ($\bm{A^*}$) is then given by
    \begin{eqnarray}
        \bm{A^*} = \bm{P} \bm{A'} \bm{P}^T  \;\;\;\; \text{where} \;\;\; \bm{P} = \left[ \begin{array}{ccc}
        cos \; \theta & ~sin \; \theta & ~0\\
        -sin \; \theta & ~cos \; \theta & ~0\\
        0 & ~0 & ~1
        \end{array} \right] 
        \label{eq:rot2}
    \end{eqnarray}
    is also a proper rotation matrix.  
\end{enumerate}
These steps result in the VGT in a rotated coordinate frame such that it is of the form
\begin{eqnarray}
    \bm{A^*} =
    \left[ \begin{array}{ccc}
    \lambda_{cr} & -\phi & 0 \\
    \phi + s_3 & \lambda_{cr} & 0 \\
    s_2 & s_1 & \lambda_r \end{array} \right] 
\end{eqnarray}
The VGT is then decomposed into the following constituent tensors,
\begin{equation*}
    \bm{A^*} = \bm{N} + \bm{H} + \bm{R}  \;\;\;\; \text{where} 
\end{equation*}
\begin{eqnarray}
    \bm{N} = \left[
    \begin{array}{ccc}
    \lambda_{cr} & 0 & 0\\
    0 & \lambda_{cr} & 0\\
    0 & 0 & \lambda_{r}
    \end{array}\right],
    \;\;\;
    \bm{H} = \left[
    \begin{array}{ccc}
    0 & 0 & 0\\
    s_3 & 0 & 0\\
    s_2 & s_1 & 0
    \end{array}\right]
    \;\;\;
    \bm{R} = \left[
    \begin{array}{ccc}
    0 & -\phi & 0\\
    \phi & 0 & 0\\
    0 & 0 & 0
    \end{array}\right]
\label{eq:NHR1}
\end{eqnarray}
We define $\bm{N}$ as the normal-strain-rate tensor, which is a diagonal tensor containing the real parts of eigenvalues of VGT. Here, $\lambda_{cr}$ is the real part of the complex conjugate eigenvalues of $\bm{A}$ and $\lambda_r$ is the only real eigenvalue of $\bm{A}$.
The tensor $\bm{N}$ represents the normal compression and expansion of the fluid element along different directions. It must be noted that the volume of the fluid element is preserved. 
Incompressibility imposes the following condition:
\begin{equation}
    \lambda_r = - 2 \lambda_{cr}
    \label{eq:incomp1}
\end{equation}
The tensor $\bm{H}$ represents the pure shearing deformation of the fluid element. The shear tensor is a lower triangular tensor with three elements - $s_1,s_2$ and $s_3$ - that constitute the transverse gradients of velocity components.
The $s_3$ component represents shearing within the plane of rigid-body-rotation and is non-negative by definition \cite{gao2019rortex}.
Moreover, the eigenvalues of such a tensor are zero.
Finally, the rigid-body-rotation tensor $\bm{R}$ is an anti-symmetric tensor depending on only one unknown, $\phi$.
The tensor represents pure-rotation of the fluid element with an angular velocity given by $\phi$.
In fact, the rotational strength is defined as twice the value of this angular velocity of the fluid element ($\tilde R \equiv 2\phi$) and thus, the Rortex vector is given by \cite{gao2019rortex},
\begin{equation}
    \bm{\vec{R}} = \tilde R \bm{\vec{r}} = 2 \phi \bm{\vec{r}} 
\end{equation}
The rigid-body-rotation tensor has one zero and two purely imaginary eigenvalues ($\pm i \phi$).
Note that this additive triple decomposition of the VGT is in the principal frame of the normal-strain-rate tensor.

\subsubsection{\label{sec:Trip1b} Non-rotational case}

In this case, we use Schur decomposition to segregate the normal-strain-rate tensor from the shear tensor, as the rigid-body-rotation is identically zero. 
In a previous study, \citet{keylock2017schur} segregated the effect of shear from the VGT by performing complex Schur decomposition, which however results in complex component tensors.
In the non-rotational case, since $\bm{A}$ contains only real eigenvalues, it can be transformed into an upper triangular tensor by real Schur decomposition:
\begin{equation}
    \bm{A^\dagger} = \bm{Q}^{*T} \bm{A} \bm{Q}^* 
    \label{eq:schur1}
\end{equation}
Here, $\bm{A^\dagger}$ is an upper triangular tensor which can now be decomposed into a diagonal (normal) tensor and a strictly upper-triangular (non-normal) tensor. $\bm{Q}^*$ is an orthogonal matrix responsible for coordinate transformation of the VGT from $\bm{A}$ to $\bm{A^\dagger}$. 
To be consistent with the lower triangular form of the shear tensor obtained in the triple decomposition method for the rotational case, we perform real Schur decomposition of the VGT transpose ($\bm{A}^T$) 
\begin{equation}
    \bm{A}^{**} = \bm{Q}^T \bm{A}^T \bm{Q}
    \label{eq:schur2}
\end{equation}
Then, the transpose of the resulting tensor yields
\begin{equation}
    \bm{A^*} = \bm{A}^{**T} = \bm{Q}^T \bm{A} \bm{Q}
\end{equation}
where $\bm{A^*}$ is the Schur decomposition of $\bm{A}$ in lower triangular form. 
Now, the VGT can be decomposed into the following normal and non-normal tensors
\begin{equation*}
    \bm{A^*} = \bm{N} + \bm{H} \;\;\; \text{where}
\end{equation*}
\begin{eqnarray}
    \bm{N} = \left[
    \begin{array}{ccc}
    \lambda_1 & 0 & 0\\
    0 & \lambda_2 & 0\\
    0 & 0 & \lambda_3
    \end{array}\right] \;,\;\;
    \bm{H} = 
    \left[
    \begin{array}{ccc}
    0 & 0 & 0\\
    s_3 & 0 & 0\\
    s_2 & s_1 & 0
    \end{array}\right]
\label{eq:NHR2}
\end{eqnarray}
Here, $\bm{N}$ is the normal tensor, i.e. a diagonal tensor containing the eigenvalues of $\bm{A}$.
It is therefore referred to as the normal-strain-rate tensor and reflects the compression and expansion that the fluid element undergoes. 
In order to ensure that the VGT decomposition is unique, the ordering of diagonal elements of the normal-strain-rate tensor is fixed to $\lambda_1 \geq \lambda_2 \geq \lambda_3$. 
This tensor consists of only two unknowns due to the incompressibility condition,
\begin{equation}
    \lambda_1 + \lambda_2 + \lambda_3 = 0 
    \label{eq:incomp2}
\end{equation}
The lower triangular tensor $\bm{H}$ is the non-normal tensor containing information about the eigenvectors of $\bm{A}$. 
This tensor consists of three independent elements representing shearing of the fluid element in three orthogonal planes and is therefore, referred to as the shear tensor. 

The VGT decomposition in both rotational and non-rotational cases are considered in the principal frame of $\bm{N}$. 
In the remaining sections of this work, the VGT will be used in this coordinate frame for convenience and will be represented by $\bm{A}$.
The results presented in this study are frame invariant and do not depend on the coordinate frame of reference.

\subsection{\label{sec:Trip2} Properties of VGT constituents}

The shear tensor can be further divided into symmetric ($\bm{H_S}$) and anti-symmetric ($\bm{H_W}$) counterparts:
\begin{equation}
     \bm{H_S} = \frac{1}{2} (\bm{H}+\bm{H}^T) \;\; \text{and} \;\; \bm{H_W} = \frac{1}{2} (\bm{H}-\bm{H}^T) 
     \label{eq:HsHw1}
\end{equation}
The symmetric-shear tensor along with normal-strain-rate tensor recovers the strain-rate tensor while the anti-symmetric-shear tensor along with rigid-body-rotation tensor constitutes the rotation-rate or vorticity tensor, i.e.
\begin{equation}
    \bm{S} = \bm{N} + \bm{H_S} \;\;\; \text{and} \;\;\; \bm{W} = \bm{H_W} + \bm{R}
    \label{eq:HsHw2}
\end{equation}
It is evident that shear contributes to vorticity as well as strain-rate. 
In this subsection, we first describe the composition of velocity gradient magnitude based on the VGT decomposition. Next, we characterize the local streamline shape associated with each of these velocity gradient constituents in the phase plane of VGT invariants.

\subsubsection{\label{sec:Trip2b}Composition of velocity gradient magnitude}

In terms of strain-rate and rotation-rate, velocity gradient magnitude (Frobenius norm: $A^2 = A_{ij}A_{ij}$) can be written as:
\begin{equation}
    A^2 = S^2 + W^2 
    \label{eq:magn0}
\end{equation}
where $S^2=S_{ij}S_{ij}$ is strain-rate magnitude related to dissipation ($\nu S^2$) and $W^2=W_{ij}W_{ij}$ is vorticity magnitude or enstrophy. 
The triple decomposition of VGT results in the following expression for magnitude:
\begin{equation}
    A_{ij}A_{ij} = \; N_{ij}N_{ij} + H_{ij}H_{ij} + R_{ij}R_{ij} + 2R_{ij}H_{ij}
    \label{eq:magn1}
\end{equation}
which can be restated as
\begin{equation}
    A^2 = \; N^2 + H^2 + R^2 + 2RH
\label{eq:magn1a}
\end{equation}
where $N^2=N_{ij}N_{ij}$, $H^2=H_{ij}H_{ij}$ and $R^2=R_{ij}R_{ij}$ are defined as the magnitudes (Frobenius norms) of normal-strain-rate, shear and rigid-body-rotation tensors, respectively. 
For locally fluid rotational case (Eq.~(\ref{eq:NHR1},\ref{eq:incomp1})), these are of the form 
\begin{equation}
    N^2 = 6\lambda_{cr}^2 \;\;, \;\; H^2 = s_1^2+s_2^2+s_3^2 \;\;,\;\; R^2 = 2\phi^2
    \label{eq:magn1b}
\end{equation}
and for the fluid non-rotational case (Eq.~(\ref{eq:NHR2},\ref{eq:incomp2})), these constituent magnitudes are of the form 
\begin{equation}
    N^2 = 2(\lambda_{1}^2+\lambda_2^2+\lambda_1\lambda_2) \;\;, \;\; H^2 = s_1^2+s_2^2+s_3^2 \;\;,\;\; R^2 = 0
    \label{eq:magn1b}
\end{equation}
The term $RH$ is defined as the correlation term between the $\bm{R}$ and $\bm{H}$ tensors. The other possible correlation terms such as $N_{ij}H_{ij}$ and $N_{ij}R_{ij}$ are identically zero since $\bm{N}$ is a diagonal matrix while both $\bm{H}$ and $\bm{R}$ are hollow matrices (all diagonal elements are zero). 
Using Eq.~(\ref{eq:NHR1}) it can be shown that the shear-rotation correlation term, which exists only if the flow is locally rotational, is given by,
\begin{equation}
    2RH = 2R_{ij}H_{ij} = 2\phi s_3
    \label{eq:magn1c}
\end{equation}
Therefore, the shear-rotation correlation term contributing to velocity gradient (VG) magnitude depends only on the rigid-body-rotation strength ($2\phi$) and the component of shear that is in the plane of rigid-body-rotation ($s_3$). Since both $\phi$ and $s_3$ are non-negative \cite{gao2019rortex} by definition, $2RH \geq 0$.

In order to measure the contribution of each component toward VG magnitude, we normalize these magnitudes by the local VG magnitude
\begin{equation}
    n^2 \equiv \frac{N^2}{A^2}\;\;, \;\; h^2 \equiv \frac{H^2}{A^2} \;\;, \;\; r^2 \equiv \frac{R^2}{A^2} \;\;,\;\; 2rh \equiv \frac{2RH}{A^2}
    \label{eq:magn2}
\end{equation}
Then, the normalized normal-strain, shear and rigid-body-rotation magnitudes have the following bounds,
\begin{equation}
    0 \leq n^2 \leq 1 \;\;,\;\; 0 \leq h^2 \leq 1 \;\;,\;\; 0 \leq r^2 \leq 1
    \label{eq:magn3}
\end{equation}
It can be shown using Eq.~(\ref{eq:NHR1}) that the normalized correlation term $2rh$ have more restricted bounds, i.e.
\begin{equation}
    0 \; \leq \; 2rh \; \leq \; \frac{1}{\sqrt{2}+1} \;\; \approx 0.41
    \label{eq:magn4}
\end{equation}
A detailed proof of the above result is included in Appendix~\ref{app:proof1}.

As shown previously in Eq.~(\ref{eq:HsHw2}), the shear tensor contributes to both $\bm{S}$ as well as $\bm{W}$ in the form of its symmetric ($\bm{H_S}$) and anti-symmetric ($\bm{H_W}$) counterparts, respectively. Since the shear tensor is a lower triangular tensor, the symmetric-shear and anti-symmetric-shear tensors are of the form
\begin{eqnarray}
    \bm{H_S} = \left[
    \begin{array}{ccc}
    0 & s_3/2 & s_2/2\\
    s_3/2 & 0 & s_1/2\\
    s_2/2 & s_1/2 & 0
    \end{array}\right] \;\; \text{and} \;\;
    \bm{H_W} = 
    \left[
    \begin{array}{ccc}
    0 & -s_3/2 & -s_2/2\\
    s_3/2 & 0 & -s_1/2\\
    s_2/2 & s_1/2 & 0
    \end{array}\right]
\label{eq:HsHw3}
\end{eqnarray}
The magnitudes of $\bm{H_S}$ and $\bm{H_W}$ are equal since,
\begin{equation}
    H_S^2 = H_W^2 = \frac{1}{2}(s_1^2+s_2^2+s_3^2)
    \label{eq:HsHw4}
\end{equation}
Therefore, the shear-magnitude $H^2$ is divided equally between strain-rate and vorticity magnitudes. From Eq.~(\ref{eq:HsHw2}) and Eq.~(\ref{eq:HsHw4}), we then obtain
\begin{equation}
    S^2 = N^2 + \frac{H^2}{2} \;\;, \;\; W^2 = R^2 + 2RH + \frac{H^2}{2}
    \label{eq:HsHw5}
\end{equation}
It may be noted that vorticity has an additional dependence on shear via the shear-rotation correlation term.

The occurrence of extreme values of $A^2$ is critical in the investigation of turbulence intermittency.
The primary goal of this study is to examine the contribution of normal-strain-rate, shear and rigid-body-rotation towards the overall VG magnitude and its relation with the pressure field.

\subsubsection{\label{sec:Trip2a} Velocity gradient composition and local streamline geometry}

In this subsection, we revisit local streamline topology and geometry using triple decomposition of VGT.
The topology of local streamlines is defined by the second and third invariants of VGT as proposed by \citet{chong1990general},
\begin{equation}
    Q_A = -\frac{1}{2}A_{ij}A_{ji} \;,\;\; R_A = -\frac{1}{3}A_{ij}A_{jk}A_{ki}
    \label{eq:QR1}
\end{equation}
In the fluid rotational case, the invariants can be expressed in terms of VGT constituents (applying Eq.~(\ref{eq:NHR1},\ref{eq:incomp1})) as follows,
\begin{equation}
    \begin{split}
    & Q_A = -3\lambda_{cr}^2 + \phi^2 + \phi s_3 = \frac{1}{2}(R^2+2RH-N^2) \\
    & R_A = 2\lambda_{cr}(\lambda_{cr}^2 + \phi^2 + \phi s_3 ) = \lambda_{cr} \bigg (\frac{N^2}{3} + R^2 +2RH \bigg )
    \end{split}
    \label{eq:QR2}
\end{equation}
It is important to note here that the invariants do not depend on the components of shear outside the plane of rigid body rotation, i.e. $s_1$ and $s_2$. 
Therefore, the topology of locally rotational streamline geometry can be expressed as a function of normal-strain-rate eigenvalues, rigid-body-rotation strength and the component of shear within the plane of rigid-body-rotation.
In the fluid non-rotational case (applying Eq.~(\ref{eq:NHR2},\ref{eq:incomp2})), the invariants are given by
\begin{equation}
    \begin{split}
    & Q_A = -(\lambda_1^2+\lambda_2^2+\lambda_1\lambda_2) = -\frac{N^2}{2} \\
    & R_A = \lambda_1\lambda_2(\lambda_1+\lambda_2) = -\lambda_1 \bigg (\frac{N^2}{2} - \lambda_1^2 \bigg )
    \end{split}
    \label{eq:QR3}
\end{equation}
In this case, both $Q_A$ and $R_A$ are functions of only the normal-strain-rate tensor and do not depend on shear at all.
Therefore, the topology of locally non-rotational streamline geometry only depends on the normal-straining.

Topology, however, only provides information about the connectivity of a geometric shape. The complete geometric shape, as shown by \citet{das2019reynolds,das2019characterization}, is better represented in the bounded phase space of normalized VGT invariants given by,
\begin{equation}
    q = \frac{Q_A}{A^2} \;,\;\; r = \frac{R_A}{(A^2)^{\frac{3}{2}}}
    \label{eq:qr}
\end{equation}
Note that these normalized invariants depend on all the shear components due to the normalization. Thus, shear might not be as critical in determining the topology of the local flow but it is important in determining its complete geometric shape.
We now examine the shape of the local streamline geometry associated with the different velocity gradient tensor components $\bm{N}$, $\bm{H}$ and $\bm{R}$ in the $q$-$r$ plane.

\begin{figure*}
\begin{tikzpicture}
\node[above right] (img) at (0,0) {\includegraphics[width=0.545\textwidth]{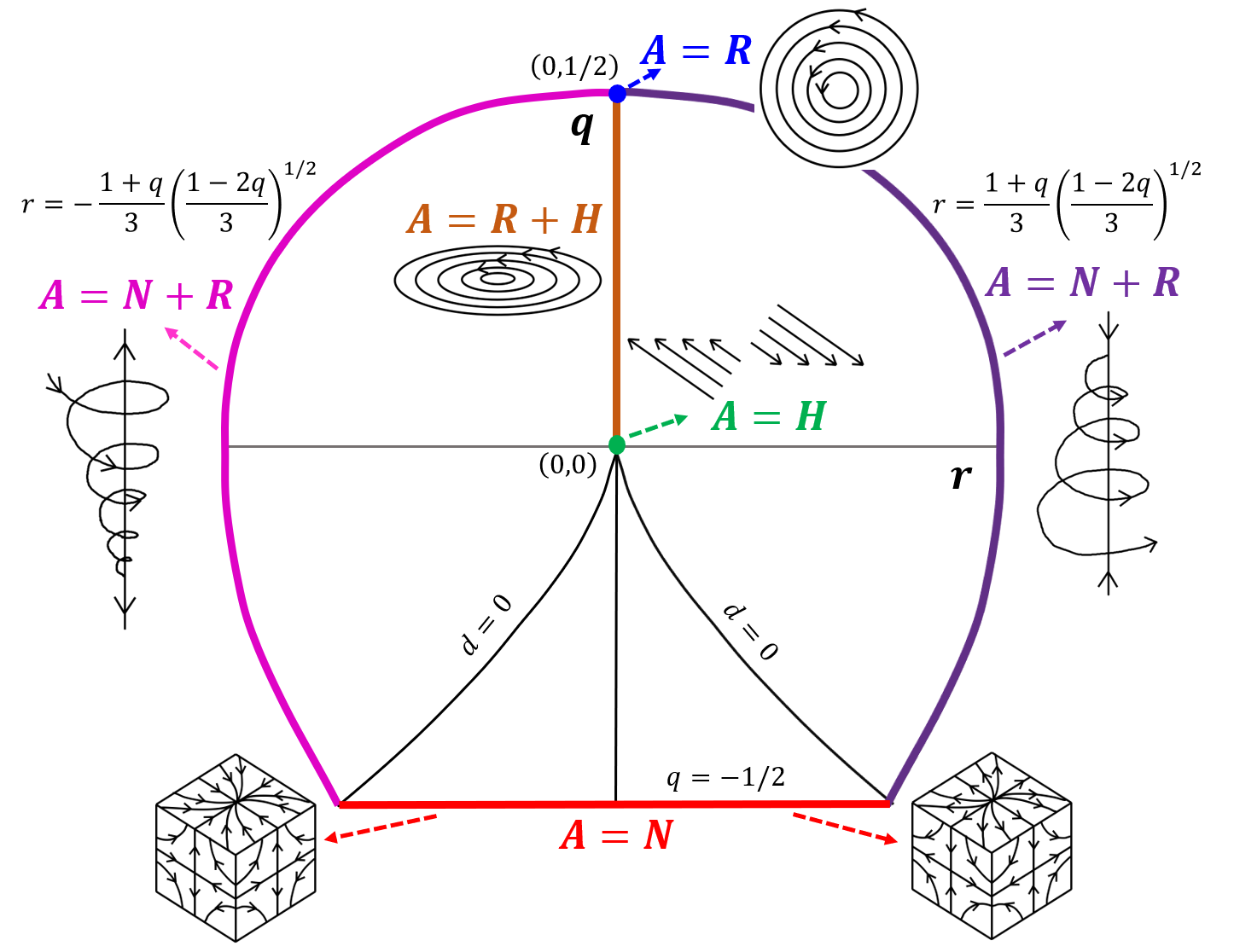}};
\node at (10pt,190pt) {(\textit{a})};
\end{tikzpicture}
\begin{tikzpicture}
\node[above right] (img) at (0,0) {\includegraphics[width=0.41\textwidth]{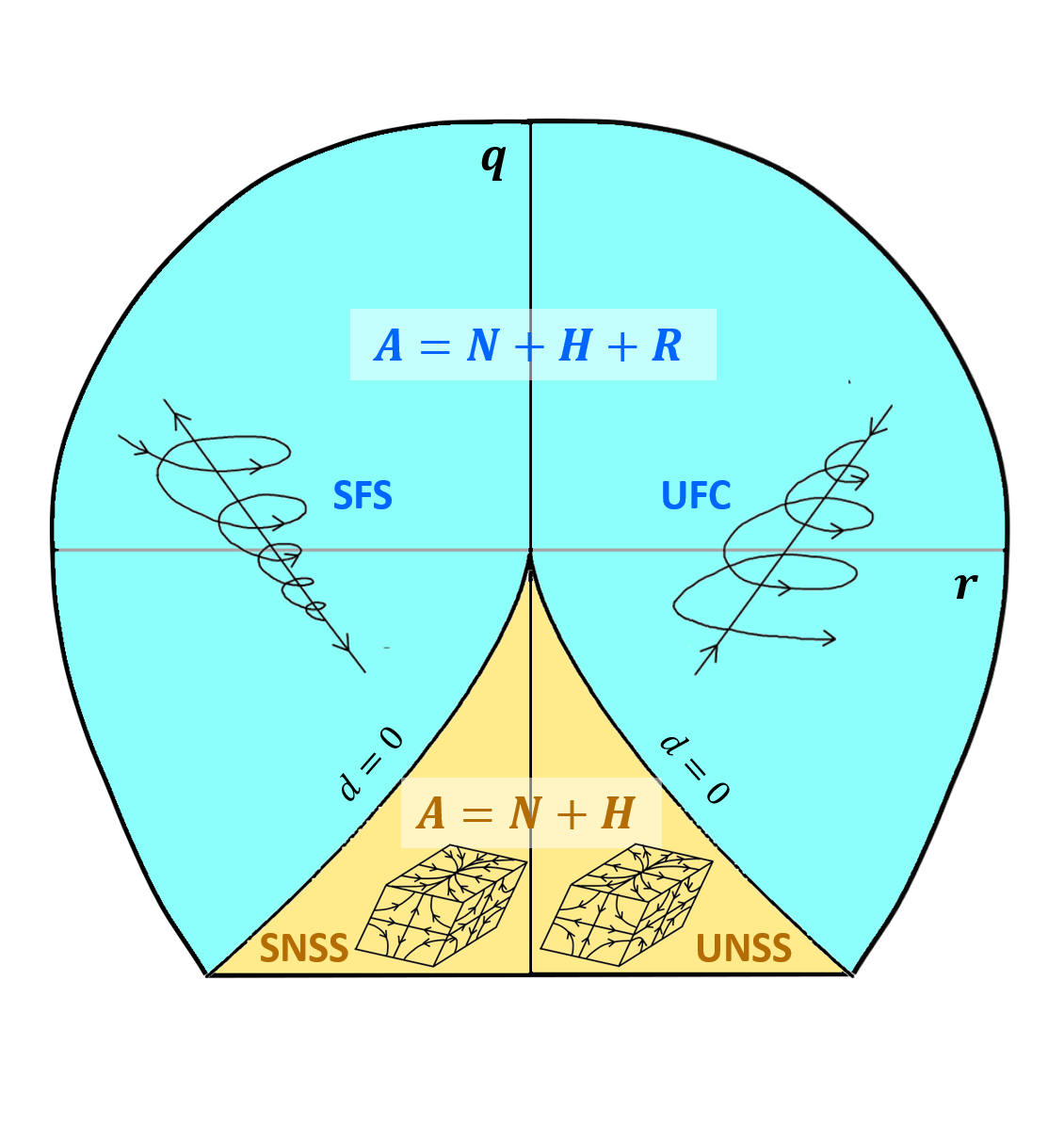}};
\node at (10pt,190pt) {(\textit{b})};
\end{tikzpicture}
\caption{\label{fig:NHR_qr} Local streamline shapes and composition of VGT in different points/regions of the $q$-$r$ plane: (a) degenerate cases (b) non-degenerate cases.}
\end{figure*}

\textit{Degenerate geometries}:  
These are the limiting cases, illustrated in Fig.~\ref{fig:NHR_qr} (a), that represent a point or a line in the $q$-$r$ plane. These degenerate shapes are discussed below:
\begin{enumerate}
    \item Pure-rotation ($\bm{A}=\bm{R}$; $\bm{N}=\bm{H}=0$): The top-most point in the plane ($r=0,q=1/2$) represents an anti-symmetric VGT with one zero and two purely imaginary eigenvalues. This results in purely rigid-body-rotation of the fluid element or locally planar circular streamlines. 
    \item Pure-shear ($\bm{A}=\bm{H}$; $\bm{N}=\bm{R}=0$): The lower triangular tensor $\bm{H}$ has zero second and third invariants by its definition. Therefore, the origin of the $q$-$r$ plane represents pure shearing of the fluid element.
    \item Normal-straining ($\bm{A}=\bm{N}$; $\bm{R}=\bm{H}=0$): The $q=-1/2$ line or bottom-most boundary of the plane constitutes the case when the VGT itself is a normal matrix. This represents pure stretching/compression of the fluid element in three orthogonal directions.
    \item Rotation and shear ($\bm{A}=\bm{R}+\bm{H}$; $\bm{N}=0$): The upper half ($q>0$) of the $r=0$ line represents VGT with zero normal-strain-rate tensor. The VGT here has one zero and two purely imaginary eigenvalues, resulting in planar closed circulating streamlines that are elliptic in shape due to the presence of shear. In the absence of shear, the streamlines are purely circular (at $q=1/2$).
    \item Normal-strain and rotation ($\bm{A}=\bm{N}+\bm{R}$; $\bm{H}=0$): The shear tensor is zero at the left and right boundaries of the plane given by
    \begin{equation}
        r = \pm \bigg ( \frac{1+q}{3} \bigg ) \bigg( {\frac{1-2q}{3}} \bigg)^{1/2}
        \label{eq:rbound}
    \end{equation}
    The left boundary represents rigid-body-rotation about the expansive normal-strain-rate eigenvector while the right boundary represents rigid-body-rotation about the compressive normal-strain-rate eigenvector, both in the absence of any shear. This implies locally stable/unstable spiraling streamlines stretching/compressing perpendicular to its focal plane.
\end{enumerate}

\textit{Non-degenerate geometries}: Fig.~\ref{fig:NHR_qr} (b) illustrates the four non-degenerate geometries covering the area of the $q$-$r$ plane and the corresponding VGT decomposition. These are discussed below:
\begin{enumerate}
    \item Rotational geometries ($\bm{A}=\bm{N}+\bm{R}+\bm{H}$): Above the zero discriminant ($d=q^3 + (27/4)r^2 =0$) lines the VGT has complex eigenvalues and the flow is locally rotational. All three constituent tensors are in general non-zero. In the stable-focus-stretching (SFS) topology, the unique eigenvalue of $\bm{N}$ (real eigenvalue of $\bm{A}$) is positive, i.e. $\lambda_r > 0$, representing stretching of the fluid element. The other two equal eigenvalues of $\bm{N}$ are negative, i.e. $\lambda_{cr} < 0$, representing convergence of the local streamlines towards a stable focus. 
    Similarly in unstable-focus-compression (UFC) topology, $\lambda_r < 0$ represents compression of the fluid element and $\lambda_{cr} >0$ represents diverging streamlines from an unstable focus.
    The non-zero eigenvalues of tensor $\bm{R}$ ($\pm i\phi$) denote the angular velocity of rigid-body-rotation of the fluid element. 
    The tensor $\bm{H}$ controls the shearing of the fluid element, resulting in varied orientations of the spiraling with respect to the direction of stretching/compression.
    The angle of alignment between the vorticity vector ($\bm{\vec{\omega}} \equiv$ dual vector of $\bm{W}$) and the eigenvectors of $\bm{S}$ can take any value. On the contrary, in this decomposition the rotation vector ($\bm{\vec{r}} \equiv$ dual vector of $\bm{R}$) is always aligned along the unique eigenvector of $\bm{N}$.
    
    \item Non-rotational geometries ($\bm{A}=\bm{N}+\bm{H}$; $\bm{R}=0$): Below the zero discriminant line, VGT has only real eigenvalues. Stable-node-saddle-saddle (SNSS) topology region represents compression in two directions and expansion in one, i.e. $\lambda_{1}>0,\lambda_{2,3}<0$,  while unstable-node-saddle-saddle (UNSS) topology implies  $\lambda_{1,2}>0,\lambda_{3}<0$. 
    However, these directions are in general oblique with respect to each other. 
    The information about the magnitude of stretching/compression is contained in $\bm{N}$ but the orientation of these directions, designated by the $\lambda_i$-eigenvectors, are contained in  $\bm{H}$.

\end{enumerate} 

In summary, pure shear occurs at the origin of the $q$-$r$ plane while normal-straining and rigid-body-rotation occur at the boundaries of the plane. The entire area inside the plane is populated in a turbulent flow field as a result of the combination of all three constituents. The distribution asymptotes to a nearly universal teardrop-like shape around the origin following the right discriminant line in fully-developed turbulent flows \cite{das2019reynolds}.

\section{\label{sec:DNS} Numerical simulation data}

Direct numerical simulation (DNS) data of incompressible forced isotropic turbulence over a range of Taylor Reynolds number ($Re_\lambda \in 1-588)$ has been used in this study to further characterize VGT behavior in terms of the new constituent tensors.
The Taylor Reynolds number is defined as
\begin{equation}
    Re_\lambda \equiv u'\lambda/\nu \;\;\;\;\; \text{based on Taylor microscale} \;\;  \lambda=(15\nu (u')^2/\epsilon)^{1/2}
\end{equation}
where $u'$ is root-mean-square (rms) velocity, $\nu$ is kinematic viscosity, and $\epsilon = 2\nu \langle S_{ij}S_{ij} \rangle $  is the mean dissipation rate.
The turbulence simulations were performed inside a $2\pi \times 2\pi \times 2\pi$ periodic box with different grid sizes listed in Table~\ref{tab:table1}.
The highest resolved wave number ($k_{max}$) normalized by the Kolmogorov length scale ($\eta$) is also listed in the table for all the DNS datasets used.
The derivatives used in this study have been calculated using spectral method.
All the DNS datasets belong to Donzis research group at Texas A$\&$M University. 
The simulation data have been used in previous works for investigating intermittency, anomalous exponents and Reynolds number scaling \cite{donzis2008,donzis2010short,gibbon2014regimes,yakhot2017emergence,yakhot2018anomalous}. 
 
\begin{table*}
\caption{\label{tab:table1}Details of forced isotropic incompressible turbulence data used}
\begin{ruledtabular}
\begin{tabular}{lccc}
 $Re_\lambda$ & Grid points & $k_{max}\eta$ & Source \\ \hline
 $~~1$ & $256^3$ & $105.6$ & \citet{yakhot2017emergence} \\
 $~~6$ & $256^3$ & $34.8$  & \citet{yakhot2017emergence} \\
 $~~9$ & $256^3$ & $26.6$ & \citet{yakhot2017emergence}\\
 $~14$ & $256^3$ & $19.87$ & \citet{yakhot2017emergence}\\
 $~18$ & $256^3$ & $15.59$ & \citet{yakhot2017emergence}\\
 $~25$ & $256^3$ & $11.51$ & \citet{yakhot2017emergence}\\
 $~35$ & $64^3$ & $1.45$ & \citet{yakhot2017emergence}\\
 $~86$ & $256^3$ & $2.83$ & \citet{donzis2008,yakhot2017emergence}\\
 $225$ & $512^3$ & $1.34$ & \citet{donzis2008,donzis2010short}\\
 $385$ & $1024^3$ & $1.41$ & \citet{donzis2008,donzis2010short}\\
 $588$ & $2048^3$ & $1.39$ & \citet{donzis2008,donzis2010short}\\
\end{tabular}
\end{ruledtabular}
\end{table*}

The work of \citet{yakhot2017emergence} demonstrated the existence of a transition Reynolds number at $Re_\lambda \sim 9$ for isotropic turbulence forced with random Gaussian forcing. 
The normalized even-order moments of velocity gradients are Gaussian below this Reynolds number and exhibit the so-called anomalous scaling above this Reynolds number. 
In addition, a recent study by \citet{das2019reynolds} show that certain VGT statistics and dynamical characteristics asymptote towards a universal nature above $Re_\lambda \approx 200$. For example, the $q$-$r$ joint probability density function (pdf) is nearly invariant for $Re_\lambda > 200$.
To better understand turbulence velocity gradient behavior as a function of \re we investigate VG composition in three ranges:
\begin{enumerate}
    \item Low Reynolds number (Gaussian regime) - $Re_\lambda \in (1,9)$
    \item Intermediate Reynolds number - $Re_\lambda \in (9,200)$
    \item High Reynolds number (Asymptotic regime) - $Re_\lambda \in (200,600)$
\end{enumerate}

\section{\label{sec:res} Velocity gradients and pressure field characterization}

\subsection{\label{sec:res1}Composition of VG magnitude}

\begin{figure}
\includegraphics[width=0.53\textwidth]{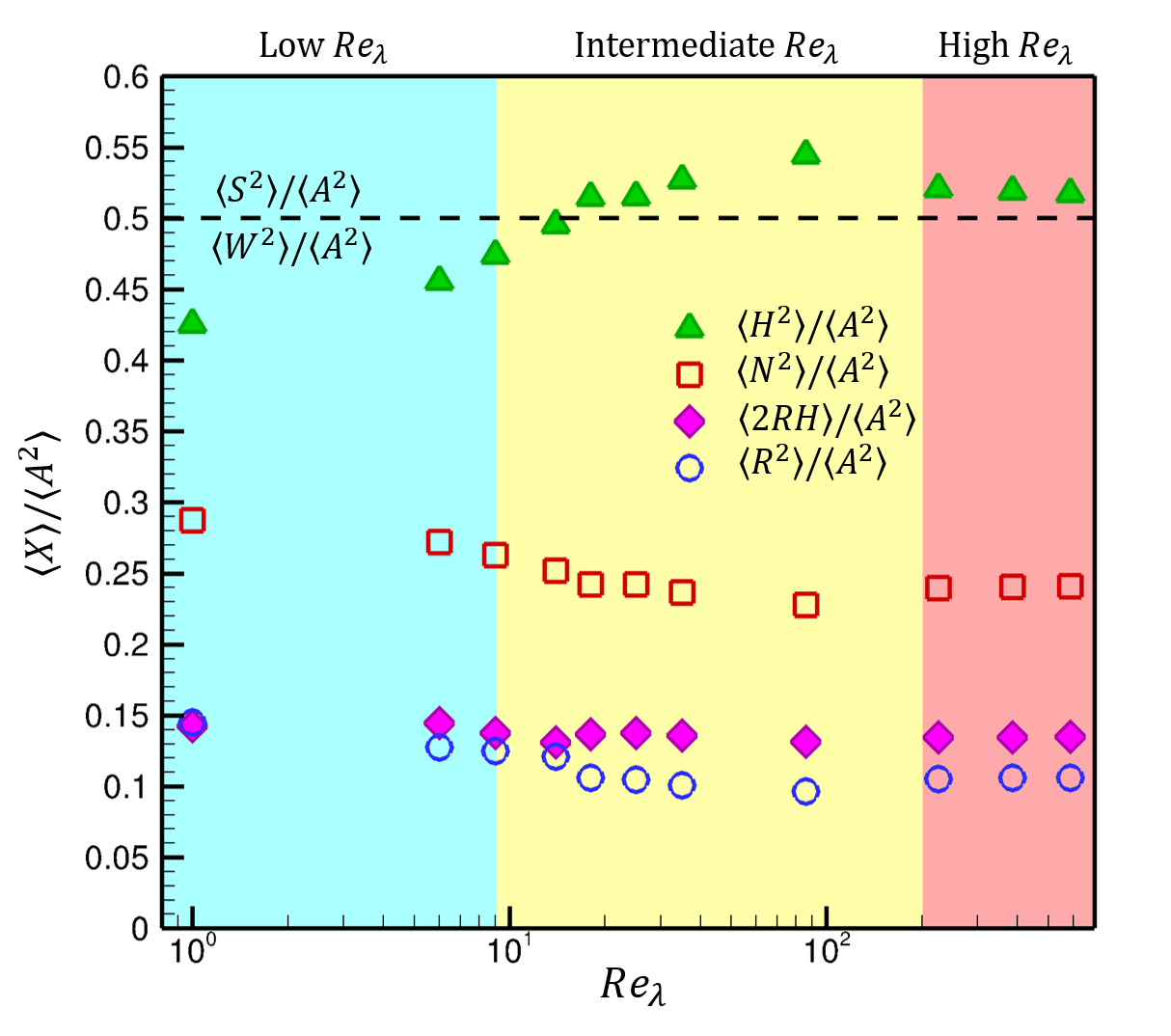}
\caption{\label{fig:GlobAvg} Volume average of $H^2$, $N^2$, $2RH$ and $R^2$ normalized by the volume average of $A^2$ in the three $Re_\lambda$ ranges (marked by different background colors). The dashed line marks the volume average of $S^2$ and $W^2$, normalized by $\langle A^2 \rangle$.}
\end{figure}

Subject to triple decomposition, the VG magnitude is composed of the following four constituents -- normal-strain-rate magnitude ($N^2$), rigid-body-rotation-rate magnitude ($R^2$), shear magnitude ($H^2$) and shear-rotation correlation term ($2RH$), as given in Eq.~(\ref{eq:magn1}). The volume-averages of these constituents normalized by the volume-average of total VG magnitude ($A^2$) is plotted in Fig.~\ref{fig:GlobAvg} as a function of Reynolds number. 
It is evident from the figure that shear is the most dominant component at all Reynolds numbers, followed by normal-strain-rate and then rigid-body-rotation. 
In the low Reynolds number range, mean shear increases with increasing \re while mean normal-strain decreases. Similar trend continues in the intermediate range of Reynolds number, where shear increases with \re to values higher than $50\%$ of the VG magnitude. 
The mean rigid-body-rotation decreases with \re in the low and intermediate ranges of Reynolds numbers. The correlation term $2RH$ is fairly independent of \re and maintains a constant value of $\langle 2RH \rangle \approx 13\%$ of $\langle A^2 \rangle $. 
Finally, the volume averages of all the constituents asymptote to distinct values in the high Reynolds number range. 
In this asymptotic regime of \rey, $\langle H^2 \rangle \approx 52 \%$, $\langle N^2 \rangle \approx 24 \%$ and $\langle R^2 \rangle \approx 11\%$ of $\langle A^2 \rangle $.

For reference, we have also plotted the VG magnitude composition in terms of ${S}_{ij}$ and $W_{ij}$. In an isotropic flow field it is well-known that $\langle S^2 \rangle/\langle A^2 \rangle = \langle W^2 \rangle/ \langle A^2 \rangle = 0.5$. 
Thus, of the $50\%$ constituted by average enstrophy or $\langle W^2 \rangle$, only about $11\%$ is directly from rigid-body-rotation. The remainder of enstrophy is constituted of contributions from shear: $\langle H_W^2 \rangle =26\%, \langle 2RH \rangle =13\%$. 
Similarly, only half of the average strain-rate magnitude or $\langle S^2 \rangle $ is constituted by normal-strain-rate; the other half is from shear. 

\begin{figure}
\includegraphics[width=0.65\textwidth]{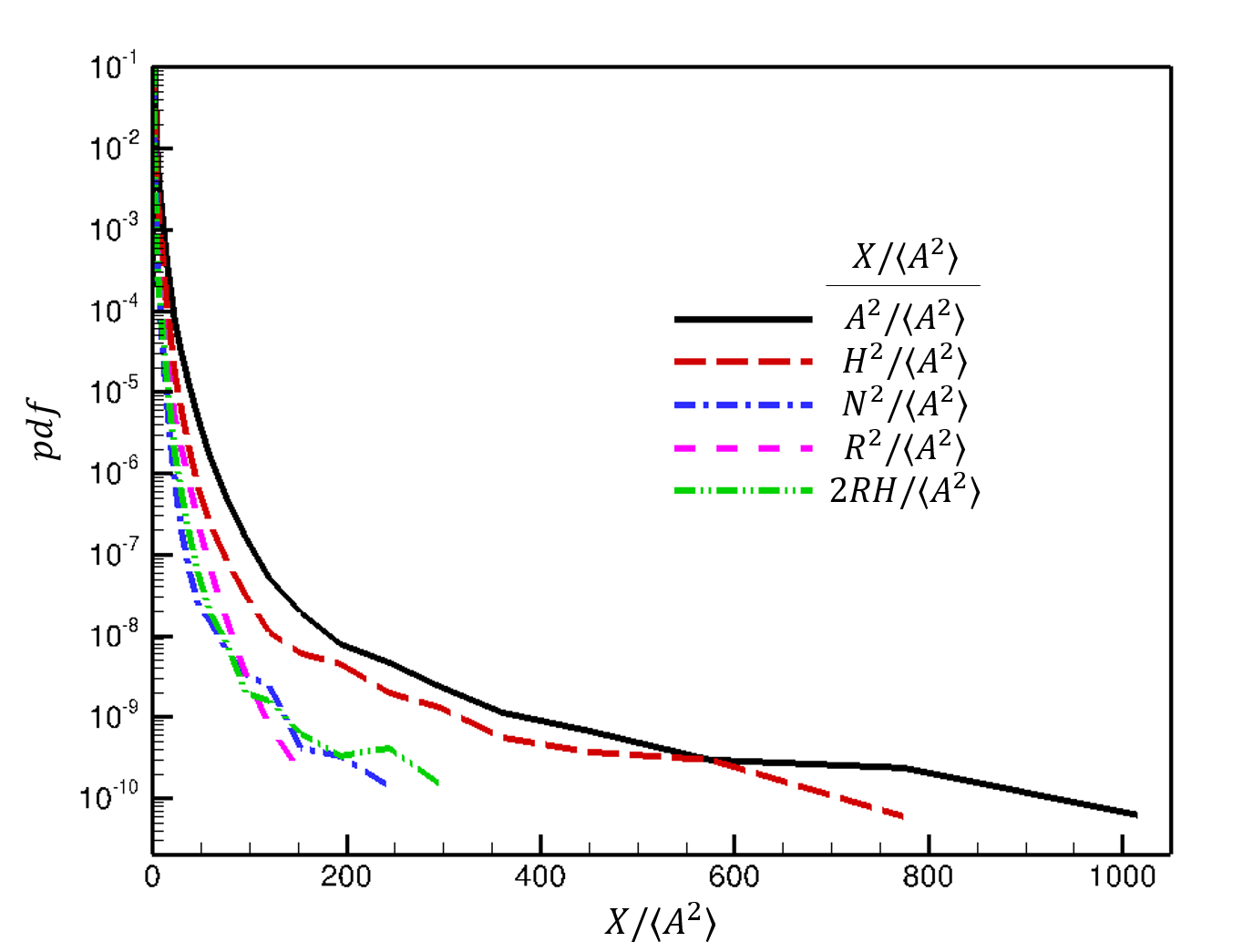}
\caption{\label{fig:pdf} Probability density function (pdf) of $A^2$, $H^2$, $N^2$, $R^2$ and $2RH$ normalized by volume-averaged VG magnitude $\langle A^2 \rangle$ in log-linear scale for $Re_\lambda=225$.}
\end{figure}

The probability density function (pdf) of $A^2$ and its composition are of much interest in the discussion of turbulence intermittency. It has been shown in several previous studies that $A^2$ exhibits a heavy-tailed pdf, characteristic of intermittency \cite{yeung1989lagrangian,yeung2006acceleration}.
It has further been shown that in the conventionally used decomposition, enstrophy exhibits a pdf with a wider tail and is more intermittent than dissipation \cite{yeung2018effects,buaria2019extreme}. 
The pdfs of VG magnitude and its triple decomposition constituents are plotted in Fig.~\ref{fig:pdf} for a high Reynolds number case (\re $=225$). 
Interestingly, the figure illustrates that shear-magnitude ($H^2$) exhibits a wider tail than all the other components. 
The pdf-tails of $N^2$, $R^2$ and $2RH$ span across smaller ranges of values than that of $H^2$. 
This is observed at all the investigated Reynolds numbers (plots not displayed) and is particularly apparent in the high \re cases.
Clearly, in this decomposition the shear-magnitude contributes most toward the heavy-tailed pdf of $A^2$.
It is further evident that it is the contribution of shear rather than rigid-body-rotation, that renders enstrophy so strongly intermittent.

\begin{figure}[h]
\includegraphics[width=0.53\textwidth]{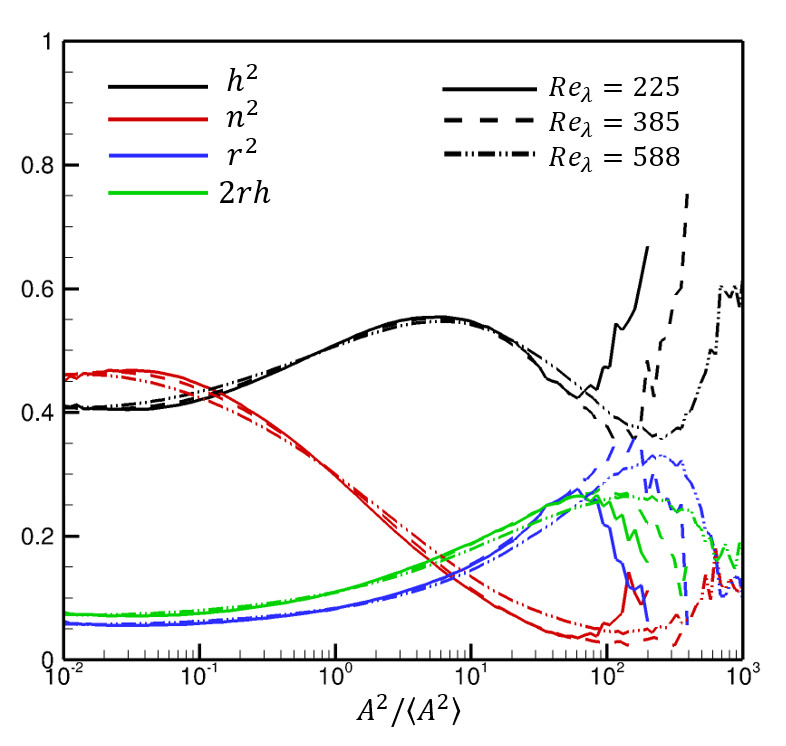}
\caption{\label{fig:condAvg} Conditional average of $h^2$, $n^2$, $r^2$ and $2rh$ as a function of $A^2/\langle A^2 \rangle$ in the high $Re_\lambda$ range.}
\end{figure}

Now we further investigate the contributions of different velocity gradient constituents at different values of $A^2$.
The conditional mean of the normalized constituents -- $n^2$, $r^2$, $h^2$ and $2rh$ (see Eq.~(\ref{eq:magn2})) are plotted as a function of VG magnitude in Fig.~\ref{fig:condAvg} for the high Reynolds number cases. 
Normal-strain-rate dominates at very low VG magnitudes ($ A^2/\langle A^2 \rangle < \mathcal{O}(0.1) $) but it declines steadily with increasing $A^2$. For a major portion of the VG magnitude range, shear is the most dominating component. 
The rigid-body-rotation magnitude and shear-rotation correlation term have very similar variation of conditional average with respect to $A^2$. 
Both $r^2$ and $2rh$ increase steadily with $A^2$, except at the extreme $A^2$ values.
There appears to be a critical value of $A^2/\langle A^2 \rangle $ in the extreme range ($\sim \mathcal{O}(10^2-10^3)$), above which $r^2$ and $2rh$ exhibit a sharp decline with $A^2$ while $h^2$ displays a steep increase. 
The contribution of $n^2$ is small in this range of $A^2/\langle A^2 \rangle$.
The critical $A^2/\langle A^2 \rangle $ value $\approx 60$ ($Re_\lambda=225$), $\approx 166$ ($Re_\lambda=385$) and $\approx 260$ ($Re_\lambda=588$), clearly increases with \rey.
The conditional average plots of all the components of VG magnitude below this critical value are nearly invariant with \rey.

This subsection demonstrates that on average shear is the dominant contributor to VG magnitude in a turbulent flow field while rigid-body-rotation magnitude contributes the least. 
It is further shown that shear-magnitude is most responsible for the heavy-tailed pdf of VG magnitude. 
Moreover, shear-magnitude increases steeply while rigid-body-rotation magnitude decreases at extreme $A^2$ values.
It is, therefore, reasonable to infer that shear dominates in regions of high intermittency.

\subsection{\label{sec:res2}Dependence of pressure field on VG constituents}

\begin{figure}[h]
\includegraphics[width=0.53\textwidth]{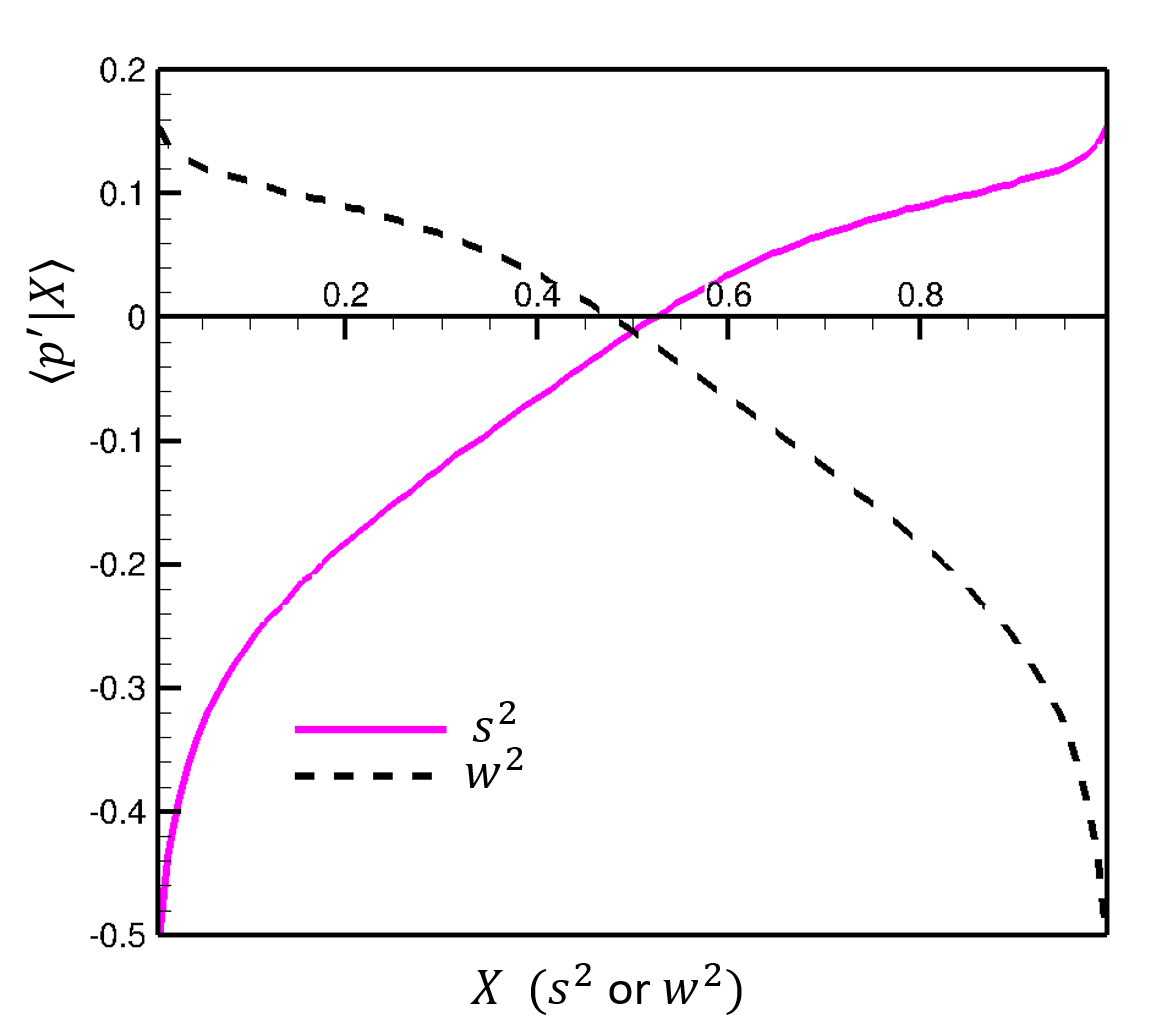}
\caption{\label{fig:p_condAvg1} Conditional average of pressure fluctuation (normalized by turbulent kinetic energy) as a function of $s^2$ and $w^2$ for $Re_\lambda=225$.}
\end{figure}

Pressure field in an incompressible turbulent flow is governed by the pressure Poisson equation wherein the source term depends on the local velocity gradients:
\begin{equation}
    \nabla^2 p' = - A_{ij}A_{ji}
    \label{eq:p1}
\end{equation}
Here $p'$ is the pressure fluctuation normalized by density.
In terms of strain-rate and vorticity tensors this equation is of the form,
\begin{equation}
    \nabla^2 p' = W_{ij}W_{ij}-S_{ij}S_{ij} = W^2 - S^2 = A^2(w^2 -s^2)
    \label{eq:p2}
\end{equation}
where, $s^2=S^2/A^2$ and $w^2=W^2/A^2$ represent the fractions of contribution of strain-rate and vorticity towards $A^2$ (Eq.~\ref{eq:magn0}).
As shown by \citet{yeung2012dissipation}, $p'$ and $\nabla^2 p'$ are negatively correlated in a homogeneous field. Therefore, we expect from Eq.~(\ref{eq:p2}) that high $s^2$ is likely associated with positive $p'$ and high $w^2$ is associated with negative $p'$.
The mean pressure fluctuation (normalized by turbulent kinetic energy $0.5\langle u'^2 \rangle$) conditioned on $s^2$ and $w^2$ are plotted in Fig.~\ref{fig:p_condAvg1} for a high \re case. The figure displays that when $s^2$ dominates in the flow, mean $p'>0$ implying a high pressure region, which is expected in a strain-dominated flow. 
On the other hand, when vorticity magnitude dominates, the local flow consists of a low pressure center.
However, note that when $w^2<1/2$ even if the flow has a significant amount of vorticity, the pressure fluctuation is likely to be positive. This reiterates the fact that vorticity does not necessarily indicate the presence of a rotating flow with a low pressure center.

Even though $\nabla^2 p'$ depends on two intermittent quantities ($S^2$ and $W^2$), it has been shown to follow Kolmogorov scaling and is essentially non-intermittent in nature \cite{iyer2019scaling}.
We now use triple decomposition of VGT to provide a plausible explanation for this.
Using Eq.~(\ref{eq:VGT1}) in the governing equation of pressure (Eq.~(\ref{eq:p1})) we obtain
\begin{equation}
    \begin{split}
    \nabla^2 p' =  - ( & R_{ij}R_{ji} + N_{ij}R_{ji} + H_{ij}R_{ji} + R_{ij}N_{ji} + N_{ij}N_{ji} \\
                & + H_{ij}N_{ji} + R_{ij}H_{ji} + N_{ij}H_{ji} + H_{ij}H_{ji})
    \end{split}
    \label{eq:p3a}
\end{equation}
Applying the properties of the tensors from Eq.~(\ref{eq:NHR1}), we can substitute the following in above equation
\begin{equation}
    R_{ij}H_{ji} = H_{ij}R_{ji} = -2RH \;\;\text{and} \;\; N_{ij}R_{ji}=R_{ij}N_{ji}=N_{ij}H_{ji}=H_{ij}N_{ji}=H_{ij}H_{ji}=0
    \label{eq:p3b}
\end{equation}
to obtain an alternate expression for source term in Laplace equation for pressure,
\begin{equation}
    \nabla^2 p' = R^2 + 2RH -N^2 = A^2(r^2 + 2rh -n^2)
    \label{eq:p3}
\end{equation}
Thus, the Laplacian of pressure does not depend on the shear-magnitude ($H^2$). The contribution of $H^2$ to the magnitudes of strain-rate and vorticity are equal, as shown in Eq.~(\ref{eq:HsHw4}), and they nullify each other in the Laplacian of pressure expression.
As shown in Section~\ref{sec:res1}, shear magnitude is the predominant contributor in attaining extreme VG magnitudes. The absence of this most intermittent VGT constituent $H^2$ renders the Laplacian of pressure significantly less intermittent than $A^2$.

It is evident in Eq.~(\ref{eq:p3}) that the sole effect of shear on pressure is through the shear-rotation correlation term that depends only on shearing in the rigid-body-rotation plane ($s_3$).
Now, if the local flow has no rigid-body-rotation at all (non-rotational), the pressure Poisson equation is given by,
\begin{equation}
    \nabla^2 p' = -N^2 = -A^2 n^2
    \label{eq:p4}
\end{equation}
In this case, the Laplacian of pressure solely depends on the magnitude of normal-strain-rate tensor.

\begin{figure}[h]
\includegraphics[width=0.53\textwidth]{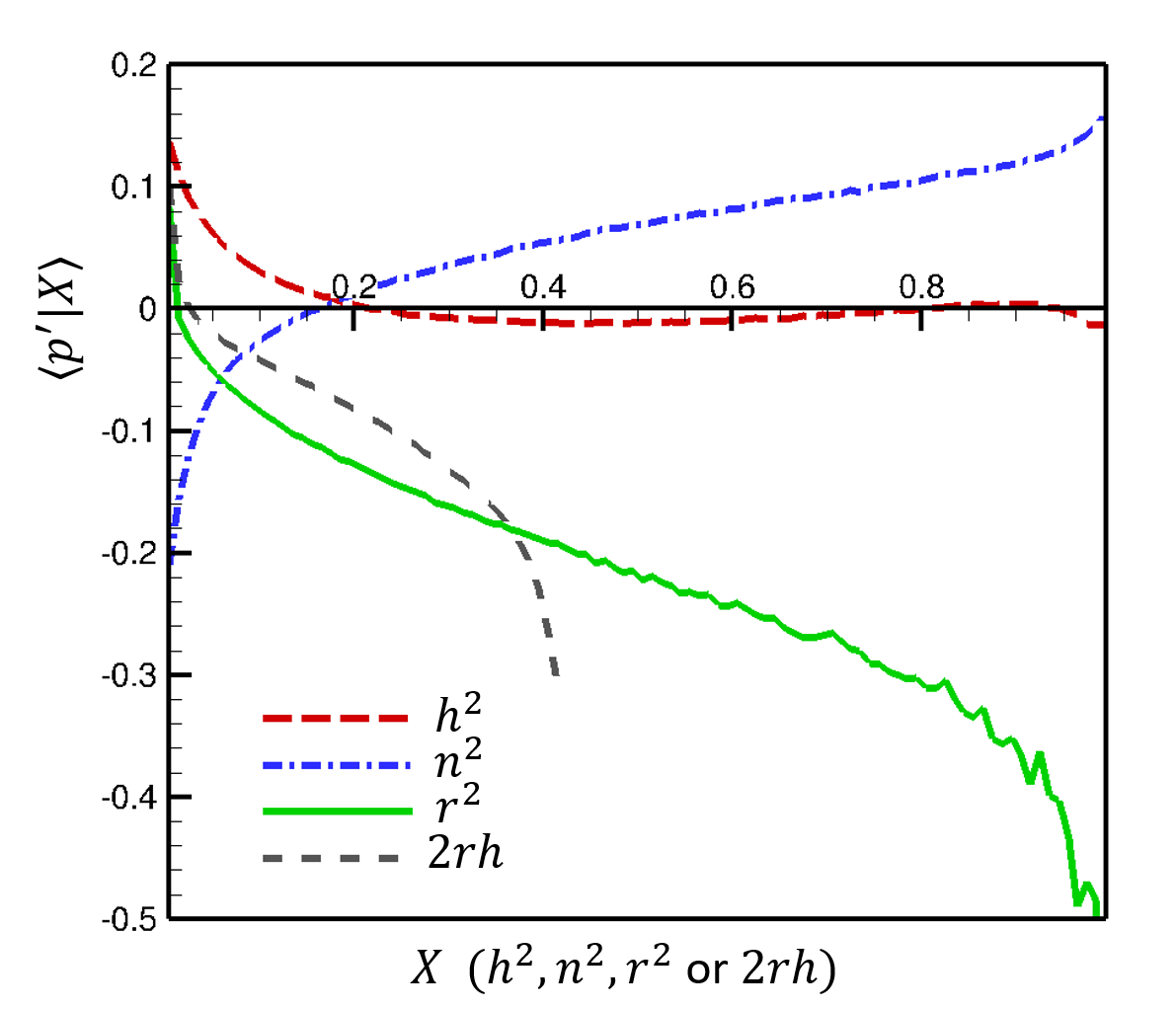}
\caption{\label{fig:p_condAvg2} Conditional average of pressure fluctuation (normalized by turbulent kinetic energy) as a function of $h^2$, $n^2$, $2rh$ and $r^2$ for $Re_\lambda=225$.}
\end{figure}

Now, we examine the mean normalized $p'$ conditioned on each of the normalized VG constituents, i.e. $n^2$, $h^2$, $r^2$ and $2rh$, in Fig.~\ref{fig:p_condAvg2}. 
It is evident that the mean pressure fluctuation is negative whenever rigid-body-rotation is present in the flow. 
Clearly, rigid-body-rotation is much more strongly correlated with low pressure regions than vorticity ($w^2$ in Fig.~\ref{fig:p_condAvg1}).
Similarly, normal-strain-rate magnitude is more strongly associated with positive $p'$ or high pressure regions than $s^2$. 
As $r^2$ increases, $\langle p'|r^2 \rangle$ becomes more negative and as $n^2$ increases, $\langle p'|n^2 \rangle$ increases. 
Shear-magnitude is mostly associated with nearly zero conditional mean pressure fluctuations. This is due to the fact that purely shearing motion does not require any pressure gradient to drive the flow and the incompressibility condition is inherently satisfied. 
The shear-rotation correlation term lies within its bounds $2rh \in (0,0.41) $ as given in Eq.~(\ref{eq:magn4}) and the pressure fluctuations conditioned on $2rh$ becomes more negative as $2rh$ increases, similar to $r^2$. 
The figures in this subsection illustrate the results for $Re_\lambda=225$; the other \re cases also display similar behavior and have not been presented separately.


\section{\label{sec:conc} Conclusions} 

The study proposes using a novel triple decomposition of velocity-gradient tensor (VGT) to revisit key small-scale features of turbulence. The VGT is partitioned into normal-strain-rate, pure-shear and rigid-body-rotation-rate tensors. Each of these tensors signifies an elementary form of deformation of a fluid element and has a specific role to play in the turbulence phenomenon.  Specifically, the decomposition permits isolating the effect of rigid-body rotation from vorticity ($W_{ij}$) and normal-strain-rate from strain-rate ($S_{ij}$). The key results and findings from this study are: 
\begin{enumerate}
    \item The various local streamline topologies and geometries can be more intuitively understood in terms of normal-strain, rigid-body-rotation and pure-shear tensors. 
    \item On average, shear is the most dominating constituent in turbulent flow fields at all Reynolds numbers while rigid-body-rotation contributes the least. 
    \item The average contribution of shear increases while that of normal-strain-rate and rigid-body-rotation-rate decreases with Reynolds number at low and intermediate \re ($< 200$). Shear-rotation correlation term does not show any \re- dependence. At high \re ($> 200$), all the average contributions are fairly invariant with Reynolds number.
    \item Shear contribution ($H^2$) is most responsible for the heavy-tailed distribution of $A^2$. Shear-magnitude shows a steep increase in contribution at extreme $A^2$ values while rotation-magnitude declines.
    \item Further, it is shown that shear, rather than rigid-body-rotation is the main cause of the strong intermittency exhibited by enstrophy.
    \item The shear-magnitude, which contributes the most in regions of high intermittency of velocity gradients, is absent in the expression for Laplacian of pressure. Thus, the pressure Laplacian does not exhibit discernible intermittency.
    \item Low-pressure regions are strongly associated with rigid-body-rotation and high-pressure regions are prevalent when normal-strain-rate dominates. Shear is associated with nearly zero pressure fluctuations.
\end{enumerate}
Overall, the study presents some novel insight into velocity-gradients and, hence, small scales of turbulence.  This work specifically highlights the key role of shear in turbulence small-scale dynamics previously attributed to strain-rate and vorticity. The new intuition developed from this triple decomposition of VGT not only leads to deeper understanding of critical turbulence phenomena, but also paves the way for improved  modeling of velocity-gradients in turbulence.

\appendix

\section{\label{app:proof1}Upper bound of normalized shear-rotation correlation term}

The VG magnitude can be expressed as follows in terms of the elements of normal-strain-rate, shear and rigid-body-rotation-rate tensors as given in Eq.~(\ref{eq:NHR1}) when the flow is locally rotational,
\begin{equation}
    A^2 = 6\lambda_{cr}^2 + 2\phi^2 + s_1^2 + s_2^2 + s_3^2 + 2 \phi s_3
    \label{eq:app1}
\end{equation}
Here, $2 \phi s_3$ is the shear-rotation correlation term which is positive by definition for rotational flow. This term is first normalized by $A^2$ such that
\begin{equation}
    2rh \; = \;\frac{2 \phi s_3}{6\lambda_{cr}^2 + 2\phi^2 + s_1^2 + s_2^2 + s_3^2 + 2 \phi s_3}
    \label{eq:app2}
\end{equation}
The next step is to obtain the maximum value that $2rh$ can attain.
Since the five variables in the above expression are independent and the denominator is a sum of non-negative terms, we can assume that $2rh$ is maximum when $\lambda_{cr}^2=s_1^2=s_2^2=0$. Note that this represents planar flow with only pure-rotation and in-plane shear.
Now, the normalized correlation term has the following upper bound
\begin{equation}
    2rh \; \leq \; \frac{2 \phi s_3}{2\phi^2 + s_3^2 + 2 \phi s_3} 
    \label{eq:app3}
\end{equation}
Dividing the numerator and denominator by $4\phi^2$, we obtain
\begin{equation}
    2rh \; \leq \; \frac{s_3/2\phi}{1/2 + (s_3/2\phi)^2 + s_3/2\phi} = \frac{x}{1/2 + x^2 + x} \equiv f(x) \;\; \text{where} \;\;\; x=s_3/2\phi
    \label{eq:app4}
\end{equation}
The maximum of function $f(x)$ occurs when
\begin{equation}
    x = \frac{1}{\sqrt{2}}
    \label{eq:app5}
\end{equation}
Therefore, the maximum value of $2rh$ is given by
\begin{equation}
    2rh \; \leq \; f \bigg (x=\frac{1}{\sqrt{2}} \bigg) \;\;\;\implies \;\;\; 2rh \leq \frac{1}{\sqrt{2}+1}
    \label{eq:app6}
\end{equation}

\bibliography{main}

\begin{thebibliography}{35}%
\makeatletter
\providecommand \@ifxundefined [1]{%
 \@ifx{#1\undefined}
}%
\providecommand \@ifnum [1]{%
 \ifnum #1\expandafter \@firstoftwo
 \else \expandafter \@secondoftwo
 \fi
}%
\providecommand \@ifx [1]{%
 \ifx #1\expandafter \@firstoftwo
 \else \expandafter \@secondoftwo
 \fi
}%
\providecommand \natexlab [1]{#1}%
\providecommand \enquote  [1]{``#1''}%
\providecommand \bibnamefont  [1]{#1}%
\providecommand \bibfnamefont [1]{#1}%
\providecommand \citenamefont [1]{#1}%
\providecommand \href@noop [0]{\@secondoftwo}%
\providecommand \href [0]{\begingroup \@sanitize@url \@href}%
\providecommand \@href[1]{\@@startlink{#1}\@@href}%
\providecommand \@@href[1]{\endgroup#1\@@endlink}%
\providecommand \@sanitize@url [0]{\catcode `\\12\catcode `\$12\catcode
  `\&12\catcode `\#12\catcode `\^12\catcode `\_12\catcode `\%12\relax}%
\providecommand \@@startlink[1]{}%
\providecommand \@@endlink[0]{}%
\providecommand \url  [0]{\begingroup\@sanitize@url \@url }%
\providecommand \@url [1]{\endgroup\@href {#1}{\urlprefix }}%
\providecommand \urlprefix  [0]{URL }%
\providecommand \Eprint [0]{\href }%
\providecommand \doibase [0]{https://doi.org/}%
\providecommand \selectlanguage [0]{\@gobble}%
\providecommand \bibinfo  [0]{\@secondoftwo}%
\providecommand \bibfield  [0]{\@secondoftwo}%
\providecommand \translation [1]{[#1]}%
\providecommand \BibitemOpen [0]{}%
\providecommand \bibitemStop [0]{}%
\providecommand \bibitemNoStop [0]{.\EOS\space}%
\providecommand \EOS [0]{\spacefactor3000\relax}%
\providecommand \BibitemShut  [1]{\csname bibitem#1\endcsname}%
\let\auto@bib@innerbib\@empty
\bibitem [{\citenamefont {Sreenivasan}\ and\ \citenamefont
  {Antonia}(1997)}]{sreenivasan1997phenomenology}%
  \BibitemOpen
  \bibfield  {author} {\bibinfo {author} {\bibfnamefont {K.~R.}\ \bibnamefont
  {Sreenivasan}}\ and\ \bibinfo {author} {\bibfnamefont {R.}~\bibnamefont
  {Antonia}},\ }\bibfield  {title} {\bibinfo {title} {The phenomenology of
  small-scale turbulence},\ }\href@noop {} {\bibfield  {journal} {\bibinfo
  {journal} {Annual review of fluid mechanics}\ }\textbf {\bibinfo {volume}
  {29}},\ \bibinfo {pages} {435} (\bibinfo {year} {1997})}\BibitemShut
  {NoStop}%
\bibitem [{\citenamefont {Yeung}\ \emph {et~al.}(2018)\citenamefont {Yeung},
  \citenamefont {Sreenivasan},\ and\ \citenamefont {Pope}}]{yeung2018effects}%
  \BibitemOpen
  \bibfield  {author} {\bibinfo {author} {\bibfnamefont {P.}~\bibnamefont
  {Yeung}}, \bibinfo {author} {\bibfnamefont {K.}~\bibnamefont {Sreenivasan}},\
  and\ \bibinfo {author} {\bibfnamefont {S.}~\bibnamefont {Pope}},\ }\bibfield
  {title} {\bibinfo {title} {Effects of finite spatial and temporal resolution
  in direct numerical simulations of incompressible isotropic turbulence},\
  }\href@noop {} {\bibfield  {journal} {\bibinfo  {journal} {Physical Review
  Fluids}\ }\textbf {\bibinfo {volume} {3}},\ \bibinfo {pages} {064603}
  (\bibinfo {year} {2018})}\BibitemShut {NoStop}%
\bibitem [{\citenamefont {Buaria}\ \emph {et~al.}(2019)\citenamefont {Buaria},
  \citenamefont {Pumir}, \citenamefont {Bodenschatz},\ and\ \citenamefont
  {Yeung}}]{buaria2019extreme}%
  \BibitemOpen
  \bibfield  {author} {\bibinfo {author} {\bibfnamefont {D.}~\bibnamefont
  {Buaria}}, \bibinfo {author} {\bibfnamefont {A.}~\bibnamefont {Pumir}},
  \bibinfo {author} {\bibfnamefont {E.}~\bibnamefont {Bodenschatz}},\ and\
  \bibinfo {author} {\bibfnamefont {P.}~\bibnamefont {Yeung}},\ }\bibfield
  {title} {\bibinfo {title} {Extreme velocity gradients in turbulent flows},\
  }\href@noop {} {\bibfield  {journal} {\bibinfo  {journal} {New Journal of
  Physics}\ }\textbf {\bibinfo {volume} {21}},\ \bibinfo {pages} {043004}
  (\bibinfo {year} {2019})}\BibitemShut {NoStop}%
\bibitem [{\citenamefont {Sanada}\ \emph {et~al.}(1991)\citenamefont {Sanada},
  \citenamefont {Ishii},\ and\ \citenamefont
  {Kuwahara}}]{sanada1991statistics}%
  \BibitemOpen
  \bibfield  {author} {\bibinfo {author} {\bibfnamefont {T.}~\bibnamefont
  {Sanada}}, \bibinfo {author} {\bibfnamefont {K.}~\bibnamefont {Ishii}},\ and\
  \bibinfo {author} {\bibfnamefont {K.}~\bibnamefont {Kuwahara}},\ }\bibfield
  {title} {\bibinfo {title} {Statistics of energy-dissipation clusters in
  three-dimensional homogeneous turbulence},\ }\href@noop {} {\bibfield
  {journal} {\bibinfo  {journal} {Progress of Theoretical Physics}\ }\textbf
  {\bibinfo {volume} {85}},\ \bibinfo {pages} {527} (\bibinfo {year}
  {1991})}\BibitemShut {NoStop}%
\bibitem [{\citenamefont {Hosokawa}\ \emph {et~al.}(1997)\citenamefont
  {Hosokawa}, \citenamefont {Oide},\ and\ \citenamefont
  {Yamamoto}}]{hosokawa1997existence}%
  \BibitemOpen
  \bibfield  {author} {\bibinfo {author} {\bibfnamefont {I.}~\bibnamefont
  {Hosokawa}}, \bibinfo {author} {\bibfnamefont {S.-i.}\ \bibnamefont {Oide}},\
  and\ \bibinfo {author} {\bibfnamefont {K.}~\bibnamefont {Yamamoto}},\
  }\bibfield  {title} {\bibinfo {title} {Existence and significance ofsoft
  worms' in isotropic turbulence},\ }\href@noop {} {\bibfield  {journal}
  {\bibinfo  {journal} {Journal of the Physical Society of Japan}\ }\textbf
  {\bibinfo {volume} {66}},\ \bibinfo {pages} {2961} (\bibinfo {year}
  {1997})}\BibitemShut {NoStop}%
\bibitem [{\citenamefont {Jimenez}\ and\ \citenamefont
  {Wray}(1998)}]{jimenez1998characteristics}%
  \BibitemOpen
  \bibfield  {author} {\bibinfo {author} {\bibfnamefont {J.}~\bibnamefont
  {Jimenez}}\ and\ \bibinfo {author} {\bibfnamefont {A.~A.}\ \bibnamefont
  {Wray}},\ }\bibfield  {title} {\bibinfo {title} {On the characteristics of
  vortex filaments in isotropic turbulence},\ }\href@noop {} {\bibfield
  {journal} {\bibinfo  {journal} {Journal of Fluid Mechanics}\ }\textbf
  {\bibinfo {volume} {373}},\ \bibinfo {pages} {255} (\bibinfo {year}
  {1998})}\BibitemShut {NoStop}%
\bibitem [{\citenamefont {Moisy}\ and\ \citenamefont
  {Jim{\'e}nez}(2004)}]{moisy2004geometry}%
  \BibitemOpen
  \bibfield  {author} {\bibinfo {author} {\bibfnamefont {F.}~\bibnamefont
  {Moisy}}\ and\ \bibinfo {author} {\bibfnamefont {J.}~\bibnamefont
  {Jim{\'e}nez}},\ }\bibfield  {title} {\bibinfo {title} {Geometry and
  clustering of intense structures in isotropic turbulence},\ }\href@noop {}
  {\bibfield  {journal} {\bibinfo  {journal} {Journal of fluid mechanics}\
  }\textbf {\bibinfo {volume} {513}},\ \bibinfo {pages} {111} (\bibinfo {year}
  {2004})}\BibitemShut {NoStop}%
\bibitem [{\citenamefont {Ashurst}\ \emph {et~al.}(1987)\citenamefont
  {Ashurst}, \citenamefont {Kerstein}, \citenamefont {Kerr},\ and\
  \citenamefont {Gibson}}]{ashurst1987alignment}%
  \BibitemOpen
  \bibfield  {author} {\bibinfo {author} {\bibfnamefont {W.~T.}\ \bibnamefont
  {Ashurst}}, \bibinfo {author} {\bibfnamefont {A.}~\bibnamefont {Kerstein}},
  \bibinfo {author} {\bibfnamefont {R.}~\bibnamefont {Kerr}},\ and\ \bibinfo
  {author} {\bibfnamefont {C.}~\bibnamefont {Gibson}},\ }\bibfield  {title}
  {\bibinfo {title} {Alignment of vorticity and scalar gradient with strain
  rate in simulated navier--stokes turbulence},\ }\href@noop {} {\bibfield
  {journal} {\bibinfo  {journal} {The Physics of fluids}\ }\textbf {\bibinfo
  {volume} {30}},\ \bibinfo {pages} {2343} (\bibinfo {year}
  {1987})}\BibitemShut {NoStop}%
\bibitem [{\citenamefont {Kerr}(1987)}]{kerr1987histograms}%
  \BibitemOpen
  \bibfield  {author} {\bibinfo {author} {\bibfnamefont {R.~M.}\ \bibnamefont
  {Kerr}},\ }\bibfield  {title} {\bibinfo {title} {Histograms of helicity and
  strain in numerical turbulence},\ }\href@noop {} {\bibfield  {journal}
  {\bibinfo  {journal} {Physical review letters}\ }\textbf {\bibinfo {volume}
  {59}},\ \bibinfo {pages} {783} (\bibinfo {year} {1987})}\BibitemShut
  {NoStop}%
\bibitem [{\citenamefont {L{\"u}thi}\ \emph {et~al.}(2009)\citenamefont
  {L{\"u}thi}, \citenamefont {Holzner},\ and\ \citenamefont
  {Tsinober}}]{luthi2009expanding}%
  \BibitemOpen
  \bibfield  {author} {\bibinfo {author} {\bibfnamefont {B.}~\bibnamefont
  {L{\"u}thi}}, \bibinfo {author} {\bibfnamefont {M.}~\bibnamefont {Holzner}},\
  and\ \bibinfo {author} {\bibfnamefont {A.}~\bibnamefont {Tsinober}},\
  }\bibfield  {title} {\bibinfo {title} {Expanding the q--r space to three
  dimensions},\ }\href@noop {} {\bibfield  {journal} {\bibinfo  {journal}
  {Journal of Fluid Mechanics}\ }\textbf {\bibinfo {volume} {641}},\ \bibinfo
  {pages} {497} (\bibinfo {year} {2009})}\BibitemShut {NoStop}%
\bibitem [{\citenamefont {Kol{\'a}{\v{r}}}(2007)}]{kolavr2007vortex}%
  \BibitemOpen
  \bibfield  {author} {\bibinfo {author} {\bibfnamefont {V.}~\bibnamefont
  {Kol{\'a}{\v{r}}}},\ }\bibfield  {title} {\bibinfo {title} {Vortex
  identification: New requirements and limitations},\ }\href@noop {} {\bibfield
   {journal} {\bibinfo  {journal} {International journal of heat and fluid
  flow}\ }\textbf {\bibinfo {volume} {28}},\ \bibinfo {pages} {638} (\bibinfo
  {year} {2007})}\BibitemShut {NoStop}%
\bibitem [{\citenamefont {Gao}\ and\ \citenamefont
  {Liu}(2019)}]{gao2019rortex}%
  \BibitemOpen
  \bibfield  {author} {\bibinfo {author} {\bibfnamefont {Y.}~\bibnamefont
  {Gao}}\ and\ \bibinfo {author} {\bibfnamefont {C.}~\bibnamefont {Liu}},\
  }\bibfield  {title} {\bibinfo {title} {Rortex based velocity gradient tensor
  decomposition},\ }\href@noop {} {\bibfield  {journal} {\bibinfo  {journal}
  {Physics of Fluids}\ }\textbf {\bibinfo {volume} {31}},\ \bibinfo {pages}
  {011704} (\bibinfo {year} {2019})}\BibitemShut {NoStop}%
\bibitem [{\citenamefont {Nagata}\ \emph {et~al.}(2019)\citenamefont {Nagata},
  \citenamefont {Watanabe}, \citenamefont {Nagata},\ and\ \citenamefont
  {da~Silva}}]{nagata2019triple}%
  \BibitemOpen
  \bibfield  {author} {\bibinfo {author} {\bibfnamefont {R.}~\bibnamefont
  {Nagata}}, \bibinfo {author} {\bibfnamefont {T.}~\bibnamefont {Watanabe}},
  \bibinfo {author} {\bibfnamefont {K.}~\bibnamefont {Nagata}},\ and\ \bibinfo
  {author} {\bibfnamefont {C.~B.}\ \bibnamefont {da~Silva}},\ }\bibfield
  {title} {\bibinfo {title} {Triple decomposition of velocity gradient tensor
  in homogeneous isotropic turbulence},\ }\href@noop {} {\bibfield  {journal}
  {\bibinfo  {journal} {Computers \& Fluids}\ ,\ \bibinfo {pages} {104389}}
  (\bibinfo {year} {2019})}\BibitemShut {NoStop}%
\bibitem [{\citenamefont {Eisma}\ \emph {et~al.}(2015)\citenamefont {Eisma},
  \citenamefont {Westerweel}, \citenamefont {Ooms},\ and\ \citenamefont
  {Elsinga}}]{eisma2015interfaces}%
  \BibitemOpen
  \bibfield  {author} {\bibinfo {author} {\bibfnamefont {J.}~\bibnamefont
  {Eisma}}, \bibinfo {author} {\bibfnamefont {J.}~\bibnamefont {Westerweel}},
  \bibinfo {author} {\bibfnamefont {G.}~\bibnamefont {Ooms}},\ and\ \bibinfo
  {author} {\bibfnamefont {G.~E.}\ \bibnamefont {Elsinga}},\ }\bibfield
  {title} {\bibinfo {title} {Interfaces and internal layers in a turbulent
  boundary layer},\ }\href@noop {} {\bibfield  {journal} {\bibinfo  {journal}
  {Physics of Fluids}\ }\textbf {\bibinfo {volume} {27}},\ \bibinfo {pages}
  {055103} (\bibinfo {year} {2015})}\BibitemShut {NoStop}%
\bibitem [{\citenamefont {{\v{S}}{\'\i}stek}\ \emph {et~al.}(2012)\citenamefont
  {{\v{S}}{\'\i}stek}, \citenamefont {Kol{\'a}{\v{r}}}, \citenamefont {Cirak},\
  and\ \citenamefont {Moses}}]{vsistek2012fluid}%
  \BibitemOpen
  \bibfield  {author} {\bibinfo {author} {\bibfnamefont {J.}~\bibnamefont
  {{\v{S}}{\'\i}stek}}, \bibinfo {author} {\bibfnamefont {V.}~\bibnamefont
  {Kol{\'a}{\v{r}}}}, \bibinfo {author} {\bibfnamefont {F.}~\bibnamefont
  {Cirak}},\ and\ \bibinfo {author} {\bibfnamefont {P.}~\bibnamefont {Moses}},\
  }\bibfield  {title} {\bibinfo {title} {Fluid-structure interaction and vortex
  identification},\ }in\ \href@noop {} {\emph {\bibinfo {booktitle}
  {Proceedings of the 18th Australasian Fluid Mechanics Conference}}}\
  (\bibinfo {organization} {Australasian Fluid Mechanics Society Australia},\
  \bibinfo {year} {2012})\BibitemShut {NoStop}%
\bibitem [{\citenamefont {Maciel}\ \emph {et~al.}(2012)\citenamefont {Maciel},
  \citenamefont {Robitaille},\ and\ \citenamefont
  {Rahgozar}}]{maciel2012method}%
  \BibitemOpen
  \bibfield  {author} {\bibinfo {author} {\bibfnamefont {Y.}~\bibnamefont
  {Maciel}}, \bibinfo {author} {\bibfnamefont {M.}~\bibnamefont {Robitaille}},\
  and\ \bibinfo {author} {\bibfnamefont {S.}~\bibnamefont {Rahgozar}},\
  }\bibfield  {title} {\bibinfo {title} {A method for characterizing
  cross-sections of vortices in turbulent flows},\ }\href@noop {} {\bibfield
  {journal} {\bibinfo  {journal} {International Journal of Heat and Fluid
  Flow}\ }\textbf {\bibinfo {volume} {37}},\ \bibinfo {pages} {177} (\bibinfo
  {year} {2012})}\BibitemShut {NoStop}%
\bibitem [{\citenamefont {Tian}\ \emph {et~al.}(2018)\citenamefont {Tian},
  \citenamefont {Gao}, \citenamefont {Dong},\ and\ \citenamefont
  {Liu}}]{tian2018definitions}%
  \BibitemOpen
  \bibfield  {author} {\bibinfo {author} {\bibfnamefont {S.}~\bibnamefont
  {Tian}}, \bibinfo {author} {\bibfnamefont {Y.}~\bibnamefont {Gao}}, \bibinfo
  {author} {\bibfnamefont {X.}~\bibnamefont {Dong}},\ and\ \bibinfo {author}
  {\bibfnamefont {C.}~\bibnamefont {Liu}},\ }\bibfield  {title} {\bibinfo
  {title} {Definitions of vortex vector and vortex},\ }\href@noop {} {\bibfield
   {journal} {\bibinfo  {journal} {Journal of Fluid Mechanics}\ }\textbf
  {\bibinfo {volume} {849}},\ \bibinfo {pages} {312} (\bibinfo {year}
  {2018})}\BibitemShut {NoStop}%
\bibitem [{\citenamefont {Dong}\ \emph {et~al.}(2019)\citenamefont {Dong},
  \citenamefont {Gao},\ and\ \citenamefont {Liu}}]{dong2019new}%
  \BibitemOpen
  \bibfield  {author} {\bibinfo {author} {\bibfnamefont {X.}~\bibnamefont
  {Dong}}, \bibinfo {author} {\bibfnamefont {Y.}~\bibnamefont {Gao}},\ and\
  \bibinfo {author} {\bibfnamefont {C.}~\bibnamefont {Liu}},\ }\bibfield
  {title} {\bibinfo {title} {New normalized rortex/vortex identification
  method},\ }\href@noop {} {\bibfield  {journal} {\bibinfo  {journal} {Physics
  of Fluids}\ }\textbf {\bibinfo {volume} {31}},\ \bibinfo {pages} {011701}
  (\bibinfo {year} {2019})}\BibitemShut {NoStop}%
\bibitem [{\citenamefont {Li}\ \emph {et~al.}(2019)\citenamefont {Li},
  \citenamefont {Yu}, \citenamefont {Wang},\ and\ \citenamefont
  {Xu}}]{li2019heat}%
  \BibitemOpen
  \bibfield  {author} {\bibinfo {author} {\bibfnamefont {H.}~\bibnamefont
  {Li}}, \bibinfo {author} {\bibfnamefont {T.}~\bibnamefont {Yu}}, \bibinfo
  {author} {\bibfnamefont {D.}~\bibnamefont {Wang}},\ and\ \bibinfo {author}
  {\bibfnamefont {H.}~\bibnamefont {Xu}},\ }\bibfield  {title} {\bibinfo
  {title} {Heat-transfer enhancing mechanisms induced by the coherent
  structures of wall-bounded turbulence in channel with rib},\ }\href@noop {}
  {\bibfield  {journal} {\bibinfo  {journal} {International Journal of Heat and
  Mass Transfer}\ }\textbf {\bibinfo {volume} {137}},\ \bibinfo {pages} {446}
  (\bibinfo {year} {2019})}\BibitemShut {NoStop}%
\bibitem [{\citenamefont {Gui}\ \emph {et~al.}(2019)\citenamefont {Gui},
  \citenamefont {Qi}, \citenamefont {Ge}, \citenamefont {Cheng}, \citenamefont
  {Wu}, \citenamefont {Yang}, \citenamefont {Tu},\ and\ \citenamefont
  {Jiang}}]{gui2019analysis}%
  \BibitemOpen
  \bibfield  {author} {\bibinfo {author} {\bibfnamefont {N.}~\bibnamefont
  {Gui}}, \bibinfo {author} {\bibfnamefont {H.-b.}\ \bibnamefont {Qi}},
  \bibinfo {author} {\bibfnamefont {L.}~\bibnamefont {Ge}}, \bibinfo {author}
  {\bibfnamefont {P.-x.}\ \bibnamefont {Cheng}}, \bibinfo {author}
  {\bibfnamefont {H.}~\bibnamefont {Wu}}, \bibinfo {author} {\bibfnamefont
  {X.-t.}\ \bibnamefont {Yang}}, \bibinfo {author} {\bibfnamefont {J.-y.}\
  \bibnamefont {Tu}},\ and\ \bibinfo {author} {\bibfnamefont {S.-y.}\
  \bibnamefont {Jiang}},\ }\bibfield  {title} {\bibinfo {title} {Analysis and
  correlation of fluid acceleration with vorticity and liutex (rortex) in
  swirling jets},\ }\href@noop {} {\bibfield  {journal} {\bibinfo  {journal}
  {Journal of Hydrodynamics}\ ,\ \bibinfo {pages} {1}} (\bibinfo {year}
  {2019})}\BibitemShut {NoStop}%
\bibitem [{\citenamefont {Arun}\ \emph {et~al.}(2019)\citenamefont {Arun},
  \citenamefont {Sameen}, \citenamefont {Srinivasan},\ and\ \citenamefont
  {Girimaji}}]{arun2019topology}%
  \BibitemOpen
  \bibfield  {author} {\bibinfo {author} {\bibfnamefont {S.}~\bibnamefont
  {Arun}}, \bibinfo {author} {\bibfnamefont {A.}~\bibnamefont {Sameen}},
  \bibinfo {author} {\bibfnamefont {B.}~\bibnamefont {Srinivasan}},\ and\
  \bibinfo {author} {\bibfnamefont {S.}~\bibnamefont {Girimaji}},\ }\bibfield
  {title} {\bibinfo {title} {Topology-based characterization of compressibility
  effects in mixing layers},\ }\href@noop {} {\bibfield  {journal} {\bibinfo
  {journal} {Journal of Fluid Mechanics}\ }\textbf {\bibinfo {volume} {874}},\
  \bibinfo {pages} {38} (\bibinfo {year} {2019})}\BibitemShut {NoStop}%
\bibitem [{\citenamefont {Liu}\ \emph {et~al.}(2018)\citenamefont {Liu},
  \citenamefont {Gao}, \citenamefont {Tian},\ and\ \citenamefont
  {Dong}}]{liu2018rortex}%
  \BibitemOpen
  \bibfield  {author} {\bibinfo {author} {\bibfnamefont {C.}~\bibnamefont
  {Liu}}, \bibinfo {author} {\bibfnamefont {Y.}~\bibnamefont {Gao}}, \bibinfo
  {author} {\bibfnamefont {S.}~\bibnamefont {Tian}},\ and\ \bibinfo {author}
  {\bibfnamefont {X.}~\bibnamefont {Dong}},\ }\bibfield  {title} {\bibinfo
  {title} {Rortex—a new vortex vector definition and vorticity tensor and
  vector decompositions},\ }\href@noop {} {\bibfield  {journal} {\bibinfo
  {journal} {Physics of Fluids}\ }\textbf {\bibinfo {volume} {30}},\ \bibinfo
  {pages} {035103} (\bibinfo {year} {2018})}\BibitemShut {NoStop}%
\bibitem [{\citenamefont {Keylock}(2017)}]{keylock2017schur}%
  \BibitemOpen
  \bibfield  {author} {\bibinfo {author} {\bibfnamefont {C.~J.}\ \bibnamefont
  {Keylock}},\ }\bibfield  {title} {\bibinfo {title} {A schur decomposition
  reveals the richness of structure in homogeneous, isotropic turbulence as a
  consequence of localised shear},\ }\href@noop {} {\bibfield  {journal}
  {\bibinfo  {journal} {arXiv preprint arXiv:1701.02541}\ } (\bibinfo {year}
  {2017})}\BibitemShut {NoStop}%
\bibitem [{\citenamefont {Chong}\ \emph {et~al.}(1990)\citenamefont {Chong},
  \citenamefont {Perry},\ and\ \citenamefont {Cantwell}}]{chong1990general}%
  \BibitemOpen
  \bibfield  {author} {\bibinfo {author} {\bibfnamefont {M.~S.}\ \bibnamefont
  {Chong}}, \bibinfo {author} {\bibfnamefont {A.~E.}\ \bibnamefont {Perry}},\
  and\ \bibinfo {author} {\bibfnamefont {B.~J.}\ \bibnamefont {Cantwell}},\
  }\bibfield  {title} {\bibinfo {title} {A general classification of
  three-dimensional flow fields},\ }\href@noop {} {\bibfield  {journal}
  {\bibinfo  {journal} {Physics of Fluids A: Fluid Dynamics}\ }\textbf
  {\bibinfo {volume} {2}},\ \bibinfo {pages} {765} (\bibinfo {year}
  {1990})}\BibitemShut {NoStop}%
\bibitem [{\citenamefont {Das}\ and\ \citenamefont
  {Girimaji}(2019{\natexlab{a}})}]{das2019reynolds}%
  \BibitemOpen
  \bibfield  {author} {\bibinfo {author} {\bibfnamefont {R.}~\bibnamefont
  {Das}}\ and\ \bibinfo {author} {\bibfnamefont {S.~S.}\ \bibnamefont
  {Girimaji}},\ }\bibfield  {title} {\bibinfo {title} {On the reynolds number
  dependence of velocity-gradient structure and dynamics},\ }\href@noop {}
  {\bibfield  {journal} {\bibinfo  {journal} {Journal of Fluid Mechanics}\
  }\textbf {\bibinfo {volume} {861}},\ \bibinfo {pages} {163} (\bibinfo {year}
  {2019}{\natexlab{a}})}\BibitemShut {NoStop}%
\bibitem [{\citenamefont {Das}\ and\ \citenamefont
  {Girimaji}(2019{\natexlab{b}})}]{das2019characterization}%
  \BibitemOpen
  \bibfield  {author} {\bibinfo {author} {\bibfnamefont {R.}~\bibnamefont
  {Das}}\ and\ \bibinfo {author} {\bibfnamefont {S.~S.}\ \bibnamefont
  {Girimaji}},\ }\bibfield  {title} {\bibinfo {title} {Characterization of
  velocity-gradient dynamics in turbulence using local streamline geometry},\
  }\href@noop {} {\bibfield  {journal} {\bibinfo  {journal} {submitted for
  publication}\ } (\bibinfo {year} {2019}{\natexlab{b}})}\BibitemShut {NoStop}%
\bibitem [{\citenamefont {Donzis}\ \emph {et~al.}(2008)\citenamefont {Donzis},
  \citenamefont {Yeung},\ and\ \citenamefont {Sreenivasan}}]{donzis2008}%
  \BibitemOpen
  \bibfield  {author} {\bibinfo {author} {\bibfnamefont {D.}~\bibnamefont
  {Donzis}}, \bibinfo {author} {\bibfnamefont {P.}~\bibnamefont {Yeung}},\ and\
  \bibinfo {author} {\bibfnamefont {K.}~\bibnamefont {Sreenivasan}},\
  }\bibfield  {title} {\bibinfo {title} {Dissipation and enstrophy in isotropic
  turbulence: {R}esolution effects and scaling in direct numerical
  simulations},\ }\href@noop {} {\bibfield  {journal} {\bibinfo  {journal}
  {Physics of Fluids}\ }\textbf {\bibinfo {volume} {20}},\ \bibinfo {pages}
  {045108} (\bibinfo {year} {2008})}\BibitemShut {NoStop}%
\bibitem [{\citenamefont {Donzis}\ and\ \citenamefont
  {Sreenivasan}(2010)}]{donzis2010short}%
  \BibitemOpen
  \bibfield  {author} {\bibinfo {author} {\bibfnamefont {D.~A.}\ \bibnamefont
  {Donzis}}\ and\ \bibinfo {author} {\bibfnamefont {K.}~\bibnamefont
  {Sreenivasan}},\ }\bibfield  {title} {\bibinfo {title} {Short-term forecasts
  and scaling of intense events in turbulence},\ }\href@noop {} {\bibfield
  {journal} {\bibinfo  {journal} {Journal of Fluid Mechanics}\ }\textbf
  {\bibinfo {volume} {647}},\ \bibinfo {pages} {13} (\bibinfo {year}
  {2010})}\BibitemShut {NoStop}%
\bibitem [{\citenamefont {Gibbon}\ \emph {et~al.}(2014)\citenamefont {Gibbon},
  \citenamefont {Donzis}, \citenamefont {Gupta}, \citenamefont {Kerr},
  \citenamefont {Pandit},\ and\ \citenamefont {Vincenzi}}]{gibbon2014regimes}%
  \BibitemOpen
  \bibfield  {author} {\bibinfo {author} {\bibfnamefont {J.~D.}\ \bibnamefont
  {Gibbon}}, \bibinfo {author} {\bibfnamefont {D.~A.}\ \bibnamefont {Donzis}},
  \bibinfo {author} {\bibfnamefont {A.}~\bibnamefont {Gupta}}, \bibinfo
  {author} {\bibfnamefont {R.~M.}\ \bibnamefont {Kerr}}, \bibinfo {author}
  {\bibfnamefont {R.}~\bibnamefont {Pandit}},\ and\ \bibinfo {author}
  {\bibfnamefont {D.}~\bibnamefont {Vincenzi}},\ }\bibfield  {title} {\bibinfo
  {title} {Regimes of nonlinear depletion and regularity in the 3d
  navier--stokes equations},\ }\href@noop {} {\bibfield  {journal} {\bibinfo
  {journal} {Nonlinearity}\ }\textbf {\bibinfo {volume} {27}},\ \bibinfo
  {pages} {2605} (\bibinfo {year} {2014})}\BibitemShut {NoStop}%
\bibitem [{\citenamefont {Yakhot}\ and\ \citenamefont
  {Donzis}(2017)}]{yakhot2017emergence}%
  \BibitemOpen
  \bibfield  {author} {\bibinfo {author} {\bibfnamefont {V.}~\bibnamefont
  {Yakhot}}\ and\ \bibinfo {author} {\bibfnamefont {D.}~\bibnamefont
  {Donzis}},\ }\bibfield  {title} {\bibinfo {title} {Emergence of multiscaling
  in a random-force stirred fluid},\ }\href@noop {} {\bibfield  {journal}
  {\bibinfo  {journal} {Physical review letters}\ }\textbf {\bibinfo {volume}
  {119}},\ \bibinfo {pages} {044501} (\bibinfo {year} {2017})}\BibitemShut
  {NoStop}%
\bibitem [{\citenamefont {Yakhot}\ and\ \citenamefont
  {Donzis}(2018)}]{yakhot2018anomalous}%
  \BibitemOpen
  \bibfield  {author} {\bibinfo {author} {\bibfnamefont {V.}~\bibnamefont
  {Yakhot}}\ and\ \bibinfo {author} {\bibfnamefont {D.~A.}\ \bibnamefont
  {Donzis}},\ }\bibfield  {title} {\bibinfo {title} {Anomalous exponents in
  strong turbulence},\ }\href@noop {} {\bibfield  {journal} {\bibinfo
  {journal} {Physica D: Nonlinear Phenomena}\ }\textbf {\bibinfo {volume}
  {384}},\ \bibinfo {pages} {12} (\bibinfo {year} {2018})}\BibitemShut
  {NoStop}%
\bibitem [{\citenamefont {Yeung}\ and\ \citenamefont
  {Pope}(1989)}]{yeung1989lagrangian}%
  \BibitemOpen
  \bibfield  {author} {\bibinfo {author} {\bibfnamefont {P.-K.}\ \bibnamefont
  {Yeung}}\ and\ \bibinfo {author} {\bibfnamefont {S.~B.}\ \bibnamefont
  {Pope}},\ }\bibfield  {title} {\bibinfo {title} {Lagrangian statistics from
  direct numerical simulations of isotropic turbulence},\ }\href@noop {}
  {\bibfield  {journal} {\bibinfo  {journal} {Journal of Fluid Mechanics}\
  }\textbf {\bibinfo {volume} {207}},\ \bibinfo {pages} {531} (\bibinfo {year}
  {1989})}\BibitemShut {NoStop}%
\bibitem [{\citenamefont {Yeung}\ \emph {et~al.}(2006)\citenamefont {Yeung},
  \citenamefont {Pope}, \citenamefont {Lamorgese},\ and\ \citenamefont
  {Donzis}}]{yeung2006acceleration}%
  \BibitemOpen
  \bibfield  {author} {\bibinfo {author} {\bibfnamefont {P.}~\bibnamefont
  {Yeung}}, \bibinfo {author} {\bibfnamefont {S.}~\bibnamefont {Pope}},
  \bibinfo {author} {\bibfnamefont {A.}~\bibnamefont {Lamorgese}},\ and\
  \bibinfo {author} {\bibfnamefont {D.}~\bibnamefont {Donzis}},\ }\bibfield
  {title} {\bibinfo {title} {Acceleration and dissipation statistics of
  numerically simulated isotropic turbulence},\ }\href@noop {} {\bibfield
  {journal} {\bibinfo  {journal} {Physics of fluids}\ }\textbf {\bibinfo
  {volume} {18}},\ \bibinfo {pages} {065103} (\bibinfo {year}
  {2006})}\BibitemShut {NoStop}%
\bibitem [{\citenamefont {Yeung}\ \emph {et~al.}(2012)\citenamefont {Yeung},
  \citenamefont {Donzis},\ and\ \citenamefont
  {Sreenivasan}}]{yeung2012dissipation}%
  \BibitemOpen
  \bibfield  {author} {\bibinfo {author} {\bibfnamefont {P.}~\bibnamefont
  {Yeung}}, \bibinfo {author} {\bibfnamefont {D.}~\bibnamefont {Donzis}},\ and\
  \bibinfo {author} {\bibfnamefont {K.}~\bibnamefont {Sreenivasan}},\
  }\bibfield  {title} {\bibinfo {title} {Dissipation, enstrophy and pressure
  statistics in turbulence simulations at high reynolds numbers},\ }\href@noop
  {} {\bibfield  {journal} {\bibinfo  {journal} {Journal of Fluid Mechanics}\
  }\textbf {\bibinfo {volume} {700}},\ \bibinfo {pages} {5} (\bibinfo {year}
  {2012})}\BibitemShut {NoStop}%
\bibitem [{\citenamefont {Iyer}\ \emph {et~al.}(2019)\citenamefont {Iyer},
  \citenamefont {Schumacher}, \citenamefont {Sreenivasan},\ and\ \citenamefont
  {Yeung}}]{iyer2019scaling}%
  \BibitemOpen
  \bibfield  {author} {\bibinfo {author} {\bibfnamefont {K.~P.}\ \bibnamefont
  {Iyer}}, \bibinfo {author} {\bibfnamefont {J.}~\bibnamefont {Schumacher}},
  \bibinfo {author} {\bibfnamefont {K.~R.}\ \bibnamefont {Sreenivasan}},\ and\
  \bibinfo {author} {\bibfnamefont {P.}~\bibnamefont {Yeung}},\ }\bibfield
  {title} {\bibinfo {title} {Scaling of locally averaged energy dissipation and
  enstrophy density in isotropic turbulence},\ }\href@noop {} {\bibfield
  {journal} {\bibinfo  {journal} {New Journal of Physics}\ }\textbf {\bibinfo
  {volume} {21}},\ \bibinfo {pages} {033016} (\bibinfo {year}
  {2019})}\BibitemShut {NoStop}%
\end{thebibliography}%

\end{document}